\documentclass[aps,prl,reprint,groupedaddress,longbibliography]{revtex4-1}
\usepackage{epsf}
\usepackage{epsfig}
\usepackage{graphicx}
\usepackage{color}
\usepackage{siunitx}
\usepackage{epstopdf}
\usepackage{amsmath}
\begin{document}
\title {\large
Sideband Modulation by Sub-Cycle Interference}
\author{S. Eckart$^{1}$}
\email{eckart@atom.uni-frankfurt.de}
\author{D. Trabert$^{1}$}
\author{K. Fehre$^{1}$}
\author{A. Geyer$^{1}$}
\author{J. Rist$^{1}$}
\author{K. Lin$^{1,2}$}
\author{F. Trinter$^{1,3}$}
\author{L. Ph. H. Schmidt$^{1}$}
\author{M. S. Sch\"offler$^{1}$}
\author{T. Jahnke$^{1}$}
\author{M. Kunitski$^{1}$}
\author{R.~D\"{o}rner$^{1}$}
\email{doerner@atom.uni-frankfurt.de}
\affiliation{$^1$ Institut f\"ur Kernphysik, Goethe-Universit\"at, Max-von-Laue-Str. 1, 60438 Frankfurt, Germany \\
$^2$ State Key Laboratory of Precision Spectroscopy, East China Normal University, 200241, Shanghai, China \\
$^3$ Molecular Physics, Fritz-Haber-Institut der Max-Planck-Gesellschaft, Faradayweg 4-6, 14195 Berlin, Germany }
\date{\today}
\begin{abstract}
We experimentally and theoretically show that the electron energy spectra strongly depend on the relative helicity in highly intense, circularly polarized two-color laser fields which is an unexpected finding. The employed counter-rotating two-color (CRTC) fields and the co-rotating two-color (CoRTC) fields are both a superposition of circularly polarized laser pulses at a central wavelength of 390\,nm and 780 nm (intensitiy ratio $I_{390}/I_{780}\approx 250$). For the CRTC field, the measured electron energy spectrum is dominated by peaks that are spaced by 3.18\,eV (corresponds to the photon energy of light at a wavelength of 390\,nm). For the CoRTC field, we observe additional energy peaks (sidebands). Using our semi-classical, trajectory-based models, we conclude that the sideband intensity is modulated by a sub-cycle interference, which sensitively depends on the relative helicity in circularly polarized two-color fields.
\end{abstract}
\maketitle

\section{I. Introduction}
When a single atom or molecule is irradiated with a highly intense light field, it can be ionized by non-resonant absorption of more photons than necessary to overcome the binding energy \cite{Keldysh1965}. This phenomenon of above threshold ionization (ATI) \cite{voronov1966many,agostini1979p,Freeman1987,Petite_ATI_1988,Misha2005} leads to peaks in the electron energy spectrum that are spaced by the photon energy. 
When photons at a second wavelength are added to the light field, the question arises what determines the relative amount of photons that is absorbed from each of the two single colors. A trivial control parameter is the intensity ratio of the two single colors. Here we show that the relative helicity of the two single colors can be an additional, very effective control parameter.

For light at a central wavelength of 390\,nm, the ATI comb in the electron energy spectrum has a spacing of $E_{390}^{ph}=3.18$\,eV. If the light field comprises photons of a second energy $E_{780}^{ph}=1.59$\,eV in addition, the electrons can have discrete energies of:
\begin{equation}
E_{\mathrm{elec}}=N_{390}\cdot E_{390}^{ph}+N_{780}\cdot E_{780}^{ph}-U_p^{\mathrm{eff}}-I_p
\label{energycons}
\end{equation}

Here, $N_{390}$ is the number of absorbed photons with an energy of $E_{390}^{ph}$ and $N_{780}$ denotes the corresponding number of photons at an energy of $E_{780}^{ph}=0.5\cdot E_{390}^{ph}$. Equation \ref{energycons} indicates that in addition to the ionization potential $I_p$ also the ponderomotive potential $U_p^{\mathrm{eff}}$ of the two-color laser field has to be taken into account \cite{Freeman1987,Ge2019}. 
To demonstrate the helicity dependence, we choose a two-color field that is the superposition of an intense circularly polarized light field at a central wavelength of 390\,nm ($I_{390}=1.2\cdot 10^{14}$\,$W/cm^2$, peak electric field of $0.041$\,a.u.) and a much weaker circularly polarized light field at a central wavelength of 780\,nm ($I_{780}=4.7\cdot 10^{11}$\,$W/cm^2$, peak electric field of $0.0026$\,a.u.). According to energy conservation (Eq. \ref{energycons}), all possible final electron energies are spaced by $E_{780}^{ph}$ (independent of the helicities of the two sinle colors). Final electron energies, $E_{\mathrm{elec}}$, that can be reached for even (odd) values of $N_{780}$ are called ATI (sideband) peaks. This is equivalent to referring to the peaks in the electron energy spectrum which result from the photons of the dominating color (at a central wavelength of 390\,nm) as ATI peaks and those in between as sidebands.

Originally, counter-rotating two-color (CRTC) fields were introduced aiming at the production of circularly polarized high-harmonics \cite{Eichmann1995,Becker1999,fleischer2014spin}. In the past years, CRTC fields and co-rotating two-color (CoRTC) fields \cite{MancusoPRA2015_CRTC} have been used to, e.g., retrieve properties of the electronic wave function \cite{Han2018}, obtain attosecond time information \cite{Ge2019, Eckart_TheoHASE2020,DanielArXiv2020}, investigate non-adiabatic offsets in momentum space at the tunnel exit \cite{Eckart2018_Offsets}, observe sub-cycle interference \cite{Zhang2014,Richter2015,Eckart2018SubCycle,DanielArXiv2020}, and to control non-sequential double ionization \cite{Mancuso2016PRL,Eckart2016,Kang2017}. Moreover, it has been shown that the ionization pathways in multi-photon ionization are different comparing counter-rotating two-color and orthogonally polarized two-color fields \cite{Kerbstadt2017} and it was discovered that the absolute ionization probability (independent of the electron energy) can be different for CRTC fields and CoRTC fields \cite{Mancuso2017_enhancement}.

The paper is organized as follows. In Sec. II, we describe our experimental setup that allows us to produce highly intense femtosecond CRTC and CoRTC laser pulses using an interferometic optical setup. Section III presents the experimental results and compares our results with our semi-classical two-step model (SCTS model) that uses non-adiabatic corrections \cite{Eckart2018_Offsets} from the saddle-point strong field approximation (SFA, see appendix). In section IV the theoretical framework from Refs. \cite{Eckart_TheoHASE2020,DanielArXiv2020} is extended to CRTC fields (Refs. \cite{Eckart_TheoHASE2020,DanielArXiv2020} only deal with CoRTC fields) which leads to a simplified theoretical model (HASE model). Based on the HASE model, section V discusses how the interplay of sub-cycle and inter-cycle interference leads to the observed differences in sideband-intensity in the electron energy spectra comparing strong field ionization by CRTC and CoRTC fields.

\section{II. Experimental Setup}
The two-color fields are generated using a $\SI{200}{\micro\meter}$ BBO to frequency double 780\,nm laser pulses (KMLabs Dragon, 40 fs FWHM, 8 kHz) using the same optical setup as in Refs. \cite{Eckart2016,EckartNatPhys2018,Eckart2018_Offsets}. We estimate the uncertainty of the absolute intensity for $\SI{780}{\nano\meter}$ and $\SI{390}{\nano\meter}$ to be $50\%$ and $20\%$, respectively. The three-dimensional electron momentum distributions from single ionization of argon presented in this work have been measured using cold-target recoil-ion momentum spectroscopy (COLTRIMS) \cite{ullrich2003recoil,jagutzki2002multiple}. The length of the electron and ion arm was $\SI{378}{\milli\meter}$ and $\SI{67.8}{\milli\meter}$, respectively. Homogeneous electric and magnetic fields of $\SI{11.4}{\volt\per\centi\meter}$ and $8.6\,$G, respectively, guided electrons and ions towards time- and position-sensitive microchannel plate detectors with hexagonal delay-line anodes \cite{jagutzki2002multiple}. During the measurement, we switch the helicity of the laser pulse at $\SI{780}{\nano\meter}$ every 240 seconds to minimize systematic errors. The total ionization rate does not depend significantly on the relative helicity.

\section{III. Results}
\begin{figure}[ht]
\begin{center}
\epsfig{file=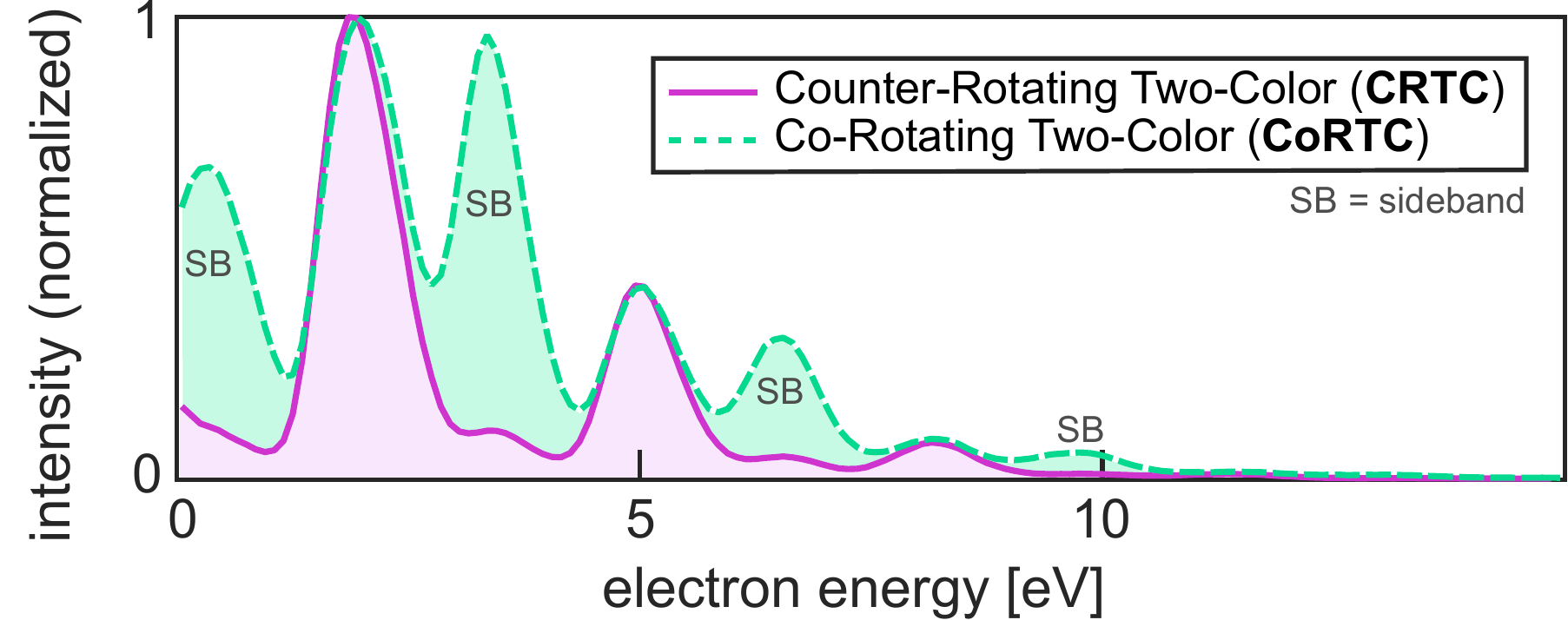, width=8.7cm}
\caption{Measured electron energy spectra depend on the relative helicity of circularly polarized two-color fields. The measured electron energy spectra upon single ionization of argon are shown for a CRTC and a CoRTC field. The sidebands are very intense for the CoRTC field and are hardly visible for the CRTC field. Sideband peaks are labeled with ``SB''.}
\label{fig_figure0label} 
\end{center}
\end{figure}

Figure \ref{fig_figure0label} shows the measured electron energy spectra upon the single ionization of argon comparing the counter-rotating two-color field (CRTC, opposite helicity of the two colors) and the co-rotating two-color field (CoRTC, same helicity of the two colors). It is evident that the CRTC field produces ATI peaks which are spaced by $E_{390}^{ph}$ and that the sidebands are hardly visible. For the CoRTC field, the sidebands are as intense as the ATI peaks. This results in an energy spectrum that is a comb with a spacing of $E_{780}^{ph}$. From the measured energy spectra it appears that there should be a straightforward explanation using selection rules or propensity rules. However, we were not able to find any explanation for our experimental results using established concepts in the strong field regime \cite{Olga2011A, Olga2011B,Fano1985}.

In this paper, we will explain the pronounced difference in sideband intensity, that is seen in Fig. \ref{fig_figure0label}, using a time-dependent perspective of wave packet creation by tunnel ionization. For a circularly polarized single-color laser field one can understand the origin of ATI peaks as a consequence of energy conservation or as a result of the periodic structure of the electron release times which acts as a grating in the time-domain and gives rise to an inter-cycle interference that leads to the periodic ATI structure in electron energy \cite{Arbo2010}. For CRTC and CoRTC fields, there is an additional sub-cycle interference which can lead to an additional modulation of the possible energy peaks according to Eq. \ref{energycons}. In this paper, we show that this sub-cycle interference is very different for CRTC and CoRTC fields and that this will elegantly explain our experimental findings.

\begin{figure}
\begin{center}
\epsfig{file=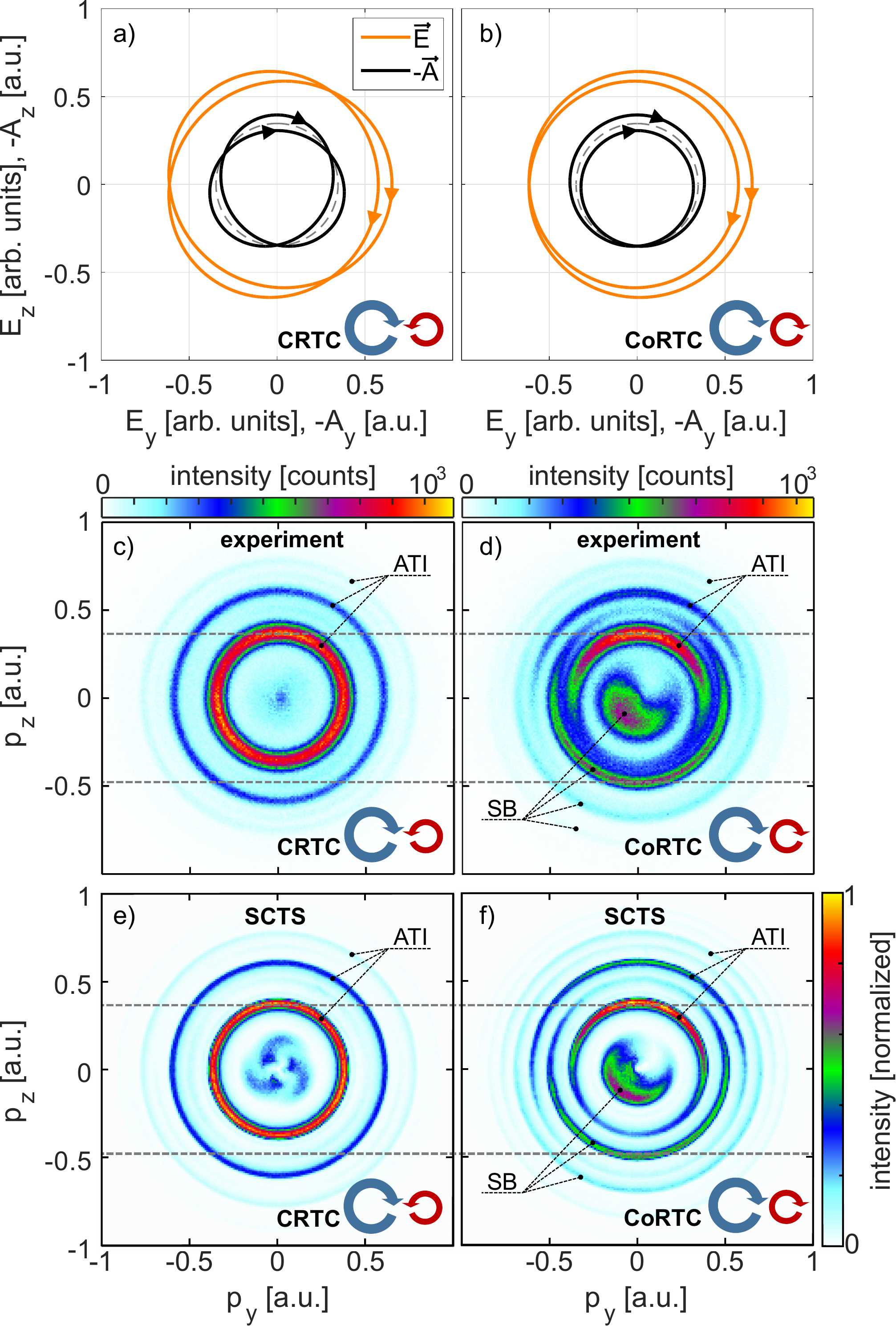, width=8.7cm}
\caption{Experimental and theoretical electron momentum distributions for counter- and co-rotating two-color fields. (a) The combined electric field $\vec{E}$ and the negative vector potential $-\vec{A}$ for the counter-rotating two-color (CRTC) field. (b) The corresponding co-rotating two-color (CoRTC) field.  The helicities of the two colors and the temporal evolution of $\vec{E}$ and $-\vec{A}$ are indicated with arrows. (c) [(d)] The measured electron momentum distribution in the plane of polarization for the laser field shown in (a) [(b)].  The electron energy spectrum shown in Fig. \ref{fig_figure0label} is based on the same data as (c) and (d). (e) [(f)] Electron momentum distribution in the plane of polarization for the CRTC [CoRTC] field that is obtained using the semi-classical two-step model (SCTS). Horizontal dashed gray lines guide the eye. Sideband peaks are labeled with ``SB''.}
\label{fig_figure1label} 
\end{center}
\end{figure}

Figures \ref{fig_figure1label}(a) and \ref{fig_figure1label}(b) depict the combined laser electric fields and the negative vector potentials that are used in the experiment. The Lissajous curves of the CRTC field and the CoRTC field are very similar. The weak laser pulse at a central wavelength of $\SI{780}{\nano\meter}$ only induces a small distortion and gives rise to a three-fold (one-fold) symmetry in the CRTC (CoRTC) field. Figure \ref{fig_figure1label}(c) shows the measured electron momentum distribution in the plane of polarization for the CRTC field. This electron momentum distribution shows almost no three-fold symmetric features and is dominated by the ATI peaks. Upon inversion of the helicity of the laser pulse at a central wavelength of $\SI{780}{\nano\meter}$, the measured electron momentum distribution changes drastically, as can be seen in Fig. \ref{fig_figure1label}(d). A strong one-fold symmetric pattern of alternating half-rings is observed. In particular, certain regions that are empty in Fig. \ref{fig_figure1label}(c) are populated in Fig. \ref{fig_figure1label}(d), giving rise to the sidebands in the electron energy spectrum (see Fig. \ref{fig_figure0label}). Fig. \ref{fig_figure1label}(e) and \ref{fig_figure1label}(f) show the results from our full theoretical model that is a semi-classical two-step (SCTS) model \cite{Shilovski2016} with tunneling probabilities from saddle-point strong field approximation (see appendix). Besides the typical overestimation of the intensity at low electron energies \cite{Eckart2018SubCycle}, the results from this SCTS model show excellent agreement with the experimental findings.

\section*{IV. A Simplified Model (HASE model)}
The alternating half-ring pattern for CoRTC fields by itself is a well-documented phenomenon \cite{Han2018,Ge2019,Eckart_TheoHASE2020,DanielArXiv2020}. The fact that the existence of sidebands depends on the relative helicity of the two single colors has not yet been described to the best of our knowledge. What is the microscopic origin of this huge difference in the energy spectra comparing CoRTC and CRTC fields? 
In the following, we show that the intensity of the ATI peaks and the sidebands are modulated by a sub-cycle interference. For CoRTC fields, this sub-cycle interference has recently been termed holographic angular streaking of electrons (HASE) and modeled using a trajectory-based, semi-classical model (HASE model) \cite{Eckart_TheoHASE2020,DanielArXiv2020}. In comparison to the full SCTS model, the HASE model neglects Coulomb interaction after tunneling, sets the momentum in the light-propagation direction to zero, uses only a two-cycle laser pulse without envelope and assumes that the tunneling probability is independent of the initial conditions. (The tunneling probability is time-independent and does not depend on the initial momentum at all.) Thus, the HASE model can be viewed to be a simplified version of the SCTS model and can be summarized as follows: We assume that the electrons are released with an initial momentum $\vec{p}_0$ that fulfills the condition 
\begin{equation}
\vec{p}_0\cdot \vec{E}(t_0)=0
\label{tunnelcondition}
\end{equation}
which indicates that $\vec{p}_0$ is perpendicular to the tunneling direction. Here $\vec{E}(t_0)$ is the laser electric field at the time the electron tunnels, $t_0$. Moreover, $\vec{p}_0$ is zero along the light propagation direction since the HASE model is a two-dimensional simulation. The final electron momentum $\vec{p}_f$ is the vectorial sum of the initial momentum upon tunneling and the negative vector potential at the electron's release time $-\vec{A}(t_0)$:
\begin{equation}
\vec{p}_f=-\vec{A}(t_0)+\vec{p}_0
\label{momcon}
\end{equation}
Since for every angle in the plane of polarization there are two different vector potentials, there are exactly two possible initial release times ($t_1$ and $t_2$) within one cycle of the two-color laser field that lead to the same final electron momentum $\vec{p}_f$. One cycle of the two-color laser field has a periodicity of $T_{780}$ (duration of one optical cycle of light at a wavelength of 780\,nm). For each final electron momentum in the plane of polarization ($p_yp_z$-plane), an optimization algorithm searches for the initial release times $t_1$ and $t_2$. 
Having found the two initial release times $t_1$ and $t_2$ within one cycle of the two-color laser field, one can consider a second, subsequent, cycle of the two-color laser field. This allows for the modeling of sub- and inter-cycle interference on the same footing \cite{Eckart_TheoHASE2020}. Therefore, the release times $t_3=t_1+T_{780}$ and $t_4=t_2+T_{780}$ are the equivalent release times in the second light cycle that lead to the same final momentum $\vec{p}_f$. Thus, within two cycles of the laser field, exactly four release times $t_n$ can be identified for each final momentum $\vec{p}_f$ (with the trajectory number $n\in\{1,2,3,4\}$). Knowing these release times and neglecting Coulomb interaction after tunneling, the semi-classical phase for these trajectories is given by \cite{Eckart_TheoHASE2020}:
\begin{equation}
\begin{aligned}
\phi_n(\vec{p}_f,t_f)&=\frac{1}{\hbar} \left( I_p t_n-\int_{t_n}^{t_f} \frac{p_y^2(t)+p_z^2(t)}{2 m_e}dt \right)\\
\end{aligned}
\label{phasemodeling}
\end{equation}

Here,  $m_e=1$\,a.u. is the electron's mass,  $\hbar=1$\,a.u. is the reduced Planck constant, and $I_p=15.76$\,eV  is the ionization potential of argon. In Eq. \ref{phasemodeling}, the first summand models the phase evolution of the electron in its bound state and the second term models the phase evolution after tunneling by the integral of the electron's kinetic energy with respect to time. Note that the choice of $t_f$ only affects the absolute phase of all four trajectories but not the relative phase of the trajectories. We always choose $t_f=t_4$ as in Ref. \cite{Eckart_TheoHASE2020}. Assuming that all four trajectories have the same probability to exist (see Ref. \cite{Eckart_TheoHASE2020} for details), the semi-classically modeled wave function at a given final electron momentum $\vec{p}_f$ is:
\begin{equation}
\begin{aligned}
\Psi(\vec{p}_f)= \sum_{n=1}^4 \exp(i\phi_n(\vec{p}_f,t_f))
\end{aligned}
\label{psisuqared}
\end{equation}
Within the HASE model, $P_{\mathrm{complete}}(\vec{p}_f)=|\Psi(\vec{p}_f)|^2$ describes intensity modulations in momentum space that are due to sub-cycle and inter-cycle interference. Trajectories are only calculated for final electron momenta $\vec{p}_f$ with an absolute value between 0.2\,a.u. and 1\,a.u. In short, the HASE model is a very intuitive and transparent model that is defined by Eq. \ref{tunnelcondition} - Eq. \ref{psisuqared}.

\begin{figure}[ht]
\begin{center}
\epsfig{file=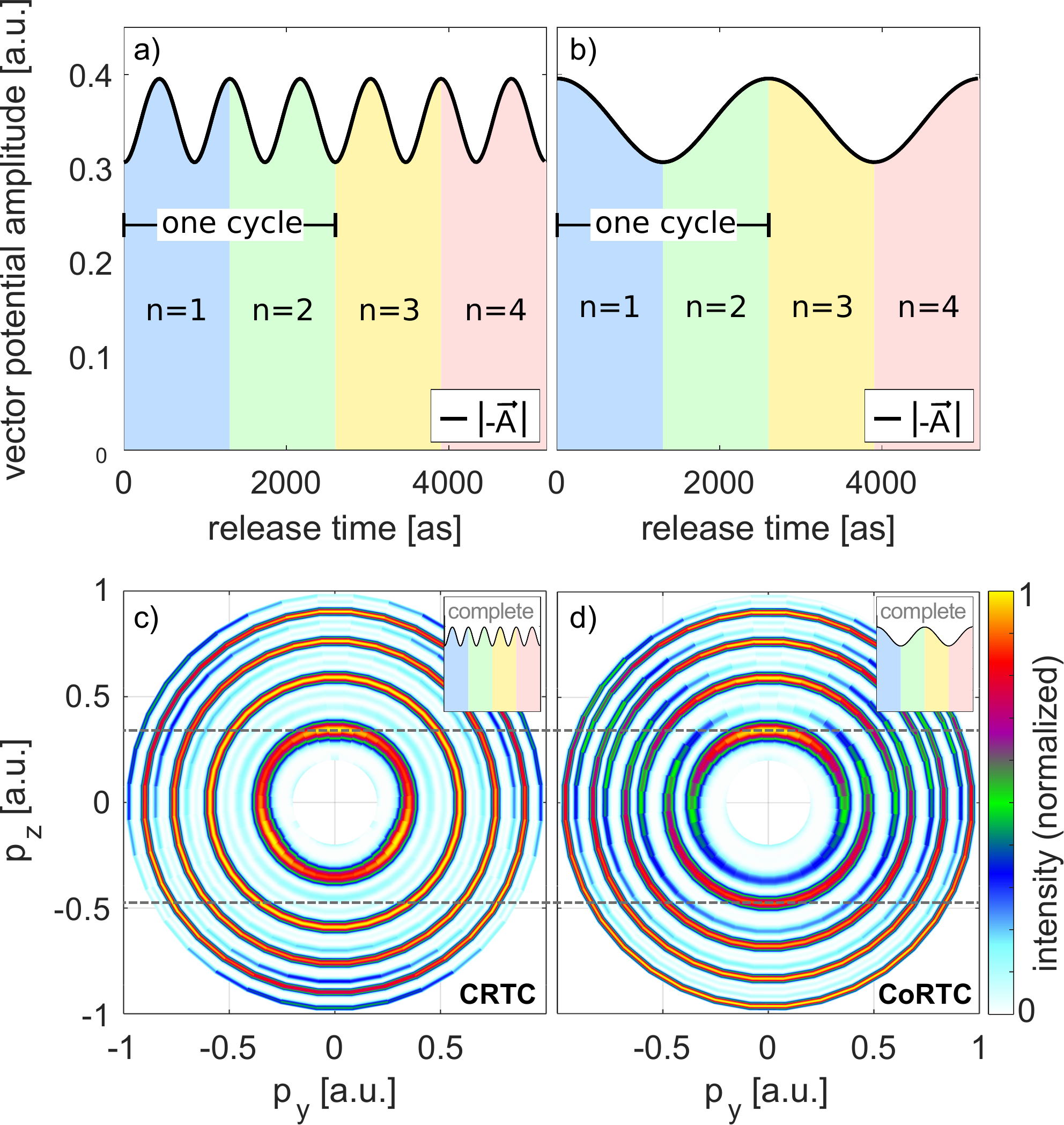, width=8.7cm}
\caption{The simplified theoretical model (HASE model) reproduces the interference patterns for counter- and co-rotating two-color fields. (a) [(b)] The absolute value of the negative vector potential $|-\vec{A}(t)|$ for two optical cycles of the CRTC [CoRTC] field. The four colored regions indicate the temporal windows of the corresponding electron release times $t_n$ and the corresponding trajectory number $n$. The duration of one cycle of the combined electric field is indicated. (c) [(d)] The semi-classically modeled intensity in final electron momentum space $P_{\mathrm{complete}}(\vec{p}_f)=\left| \sum_{n=1}^4 \exp(i\phi_n(\vec{p}_f)) \right|^2$ for the CRTC [CoRTC] field (see text related to Eq. \ref{psisuqared}). Horizontal dashed gray lines guide the eye in (c) and (d).}
\label{fig_figure2label} 
\end{center}
\end{figure}

The results from the HASE model are shown in Fig. \ref{fig_figure2label}. Figures \ref{fig_figure2label}(a) and \ref{fig_figure2label}(b) show the absolute value of the negative vector potential $|-\vec{A}(t)|$ for two optical cycles of the CRTC and the CoRTC field, respectively. The possible ranges of the release times $t_1$, $t_2$, $t_3$, and $t_4$ are colored in blue, green, yellow, and red and are labeled with the corresponding trajectory number $n$. It should be noted that for every point in final momentum space, the values of $t_n$ can differ but the allowed ranges, that are indicated with colors, do not change as a function of $\vec{p}_f$ \cite{Eckle2008,Eckart_TheoHASE2020}. Figure \ref{fig_figure2label}(c) [\ref{fig_figure2label}(d)] shows the semi-classically modeled intensity $P_{\mathrm{complete}}(\vec{p}_f)=|\Psi(\vec{p}_f)|^2$ in final electron momentum space for the CRTC [CoRTC] field. Strikingly, the semi-classical result for the CRTC field shows almost no intensity at the energies that correspond to the sidebands (Fig. \ref{fig_figure2label}(c)). The alternating half-ring structure is only observed for the CoRTC field (Fig. \ref{fig_figure2label}(d)). For the CRTC field, only a weak modulation in intensity as a function of the angle in the plane of polarization is visible. This is all in excellent agreement with the findings presented in Fig. \ref{fig_figure1label}. The fact that the HASE model succeeds to model the experimentally observed interference pattern shows that the HASE model captures the essential physics of the studied scenarios.

\section{V. Discussion}
\begin{figure}[ht]
\begin{center}
\epsfig{file=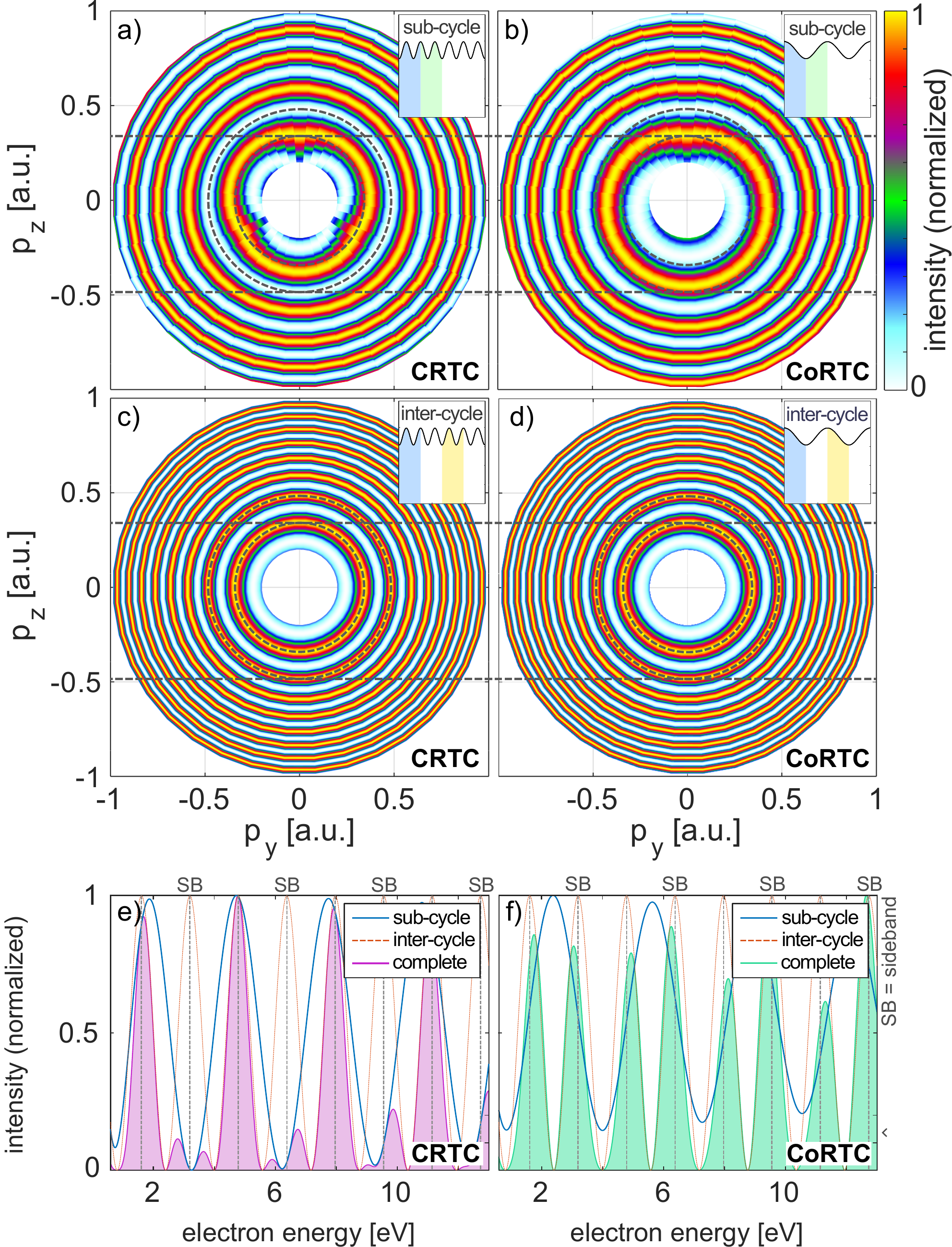, width=8.7cm}
\caption{(a) [(b)] Intensity modulation due to sub-cycle interference $P_{\mathrm{sub}}=|\exp(i\phi_1)+ \exp(i\phi_2)|^2$ for the CRTC [CoRTC] field. The possible release times of the trajectories for $\phi_1$  and $\phi_2$ are labeled with $n=1$ and $n=2$  in Fig. \ref{fig_figure2label}. (c), (d) Intensity modulation due to inter-cycle interference $P_{\mathrm{inter}}=|\exp(i\phi_1)+ \exp(i\phi_3)|^2$ (release times are labeled with $n=1$ and $n=3$ in Fig. \ref{fig_figure2label}). (e) Interference patterns from (a), (c), and Fig. \ref{fig_figure2label}(c) as a function of the electron energy for the CRTC field. (f) The same as (e) for the CoRTC field. Vertical gray dashed lines indicate the peaks of the inter-cycle interference. The dashed gray lines in (a)-(d) are the same in all four panels and guide the eye.}
\label{fig_figure3label} 
\end{center}
\end{figure}

What is the reason for the obvious difference comparing the CRTC and the CoRTC field? Since our semi-classical results only model sub-cycle and inter-cycle interference, we inspect the relative phases of the interfering semi-classical trajectories. To this end, the intensity $P_{\mathrm{sub}}=|\exp(i\phi_1)+ \exp(i\phi_2)|^2$ is  visualized for the CRTC field as a function of the final electron momentum in the plane of polarization (Fig. \ref{fig_figure3label}(a)). High intensities indicate constructive interference and therefore a phase difference $\phi_2-\phi_1$ that is close to multiples of $2\pi$. This interference pattern is due to a sub-cycle interference since the difference in the release time of the two contributing trajectories is $t_2-t_1$, which is shorter than one light cycle (see Fig. \ref{fig_figure2label}). Figure \ref{fig_figure3label}(b) shows the same for the CoRTC field. Comparison of Figs. \ref{fig_figure3label}(a) and \ref{fig_figure3label}(b) reveals that the phase difference $\phi_2-\phi_1$ depends more strongly on the angle in the plane of polarization for the CoRTC field. This can be clearly seen by comparing the interference pattern near the gray dashed circles that guide the eye in Figs. \ref{fig_figure3label}(a) and \ref{fig_figure3label}(b). In a next step, the relative phase $\phi_3-\phi_1$ is investigated. Figures \ref{fig_figure3label}(c) and  \ref{fig_figure3label}(d) visualize $P_{\mathrm{inter}}=|\exp(i\phi_1)+ \exp(i\phi_3)|^2$. Because the time difference $t_3-t_1$ is exactly as long as one light cycle, the interference is an inter-cycle interference. Figures \ref{fig_figure3label}(c) and  \ref{fig_figure3label}(d) show the same distribution which is independent of the angle in the plane of polarization. The two inner rings with high intensity in Figs. \ref{fig_figure3label}(c) and \ref{fig_figure3label}(d) are highlighted with gray dashed circles. If high intensity is observed for $P_{\mathrm{sub}}$, then $\phi_2-\phi_1$ is close to multiples of $2\pi$. And if high intensity is observed for $P_{\mathrm{inter}}$, then $\phi_3-\phi_1$ is close to multiples of $2\pi$. Thus, if high intensity is observed for $P_{\mathrm{sub}}$ and $P_{\mathrm{inter}}$, then $\phi_3-\phi_2$ is also close to multiples of $2\pi$. Further, the relative phase $\phi_4-\phi_3$ is the same as $\phi_2-\phi_1$ because of the periodicity of the light field (see Fig. \ref{fig_figure2label}). This allows to conclude, that a high intensity for $P_{\mathrm{sub}}(\vec{p}_f)$ and $P_{\mathrm{inter}}(\vec{p}_f)$ implies that all four trajectories are interfering constructively for this $\vec{p}_f$ which leads to high intensity in final electron momentum space ($P_{\mathrm{complete}}(\vec{p}_f)$). Comparison of Fig. \ref{fig_figure2label}(d) with Fig. \ref{fig_figure3label}(b) and Fig. \ref{fig_figure3label}(d) reveals that it is in fact the interplay of sub- and inter-cycle interference that determines the final electron momentum distribution $P_{\mathrm{complete}}$ and gives rise to the alternating half-ring pattern for the CoRTC field. The weak modulations for the CRTC field that are seen in Fig. \ref{fig_figure2label}(c) can be explained in full analogy.

These insights allow for a microscopic explanation of the experimental findings presented in Fig. \ref{fig_figure0label} and Fig. \ref{fig_figure1label}: 
The inter-cycle interference $P_{\mathrm{inter}}$, that is seen in Figs. \ref{fig_figure3label}(c) and  \ref{fig_figure3label}(d), reflects a comb of allowed final electron energies that is independent of the angle in the polarization plane. This is in line with the allowed electron energies according to energy conservation (see Eq. \ref{energycons}). The finding that the sideband intensity depends on the relative helicity of the two colors is explained by a sub-cycle interference that depends on the angle in the polarization plane and is very different for the CRTC and the CoRTC field. Figures \ref{fig_figure3label}(e) and \ref{fig_figure3label}(f) illustrate this in energy space showing the energy spectra of the momentum distributions $P_{\mathrm{sub}}$, $P_{\mathrm{inter}}$, and $P_{\mathrm{complete}}$. It is evident that the sub-cycle interference is very different comparing the CRTC and the CoRTC field: For the CRTC field, the sub-cycle interference is destructive at the energies that belong to the sidebands which leads to a strong suppression of sideband intensity in the corresponding final spectra. For the CoRTC field the sub-cycle interference attenuates ATI peaks and sidebands similarly but at different angle in the plane of polarization in final momentum space. In particular, it is evident from Fig. \ref{fig_figure3label}(b) that it is not possible to find an electron energy for which the sub-cycle interference is destructive for all angles in the polarization plane which explains why the curve for the sub-cycle interference pattern as a function of energy (see Fig. \ref{fig_figure3label}(f)) is bigger than zero for all electron energies.

In conclusion, our experimental results show an unexpected dependence of sideband intensity on the relative helicity in circularly polarized two-color light fields. We have used two different trajectory-based, semi-classical models that both reproduce the almost complete suppression of sideband intensity for the CRTC field. The modulation of the sideband intensity is a consequence of the differences in sub-cycle interference comparing CRTC and CoRTC fields. Our findings enable a better understanding of above-threshold ionization and are an important insight regarding the coherent control of electrons. The overall ionization probability is governed by the $2\omega$ field which is about 100 times more intense than the $\omega$ field. The emission of an entire class of electrons (sideband electrons) can be switched on and off on femtosecond timescales by employing an $\omega$ field with time-dependent polarization \cite{Corkum_1994}. This is similar to a transistor in electronics and would represent an electron emitter that is powered by a strong laser field at $2\omega$ (``transistor's collector'') and can be controlled by the helicity (which is equivalent to a relative phase for a light wave or to angular momentum for a photon) of a relatively weak laser pulse at a different frequency $\omega$ (``transistor's base'').

\begin{acknowledgments}
\section{acknowledgments}
\normalsize
This work was funded by the German Research Foundation (DFG) through Priority Programme SPP 1840 QUTIF. K.F. acknowledges support by the German Academic Scholarship Foundation. K.L. acknowledges support by the Alexander von Humboldt Foundation. S.E. acknowledges a highly appreciated talk by Dieter Bauer on 'Strong-field few-body dynamics' that was held in 2016 and which contained Eq. \ref{SFArate}.
\end{acknowledgments}

\section{Appendix: SCTS Calculations}
The semi-classical two-step (SCTS) simulation is based on the procedure described in Ref. \cite{Shilovski2016}. Instead of using Eq. 9 from Ref. \cite{Shilovski2016} we use the ionization probability from saddle-point strong field approximation (SFA) \cite{Misha2005,Milosevic2006,Popruzhenko2014}. In SFA the ionization probability for a circularly polarized light field depends on the frequency of the light field $\omega_{\mathrm{SFA}}$, the magnitude of the instantaneous electric field $E_{\mathrm{SFA}}$, the ionization potential $I_p$, the momentum in the light propagation direction $p_{x0}$ at the tunnel exit and the initial momentum at the tunnel exit which is in the plane of polarization and perpendicular to the tunnel exit $p_0$. For our SCTS simulations we use $\omega_{\mathrm{SFA}}=2 \omega$ with $\omega=0.0584$\,a.u. and hereby neglect the changes in ionization probability that are due to the subcycle changes in angular frequency. This is a good approximation because the condition $I_{780}\ll I_{390}$ is fulfilled in our experiment. Neglecting the pre-exponential factor in saddle-point SFA, the instantaneous ionization probability can be approximated by \cite{Olga2011A,Popruzhenko2014}:
\begin{equation}
\begin{aligned}
P = \exp \left(- \frac{2\cdot I_p^{\mathrm{eff}}}{\omega_{\mathrm{SFA}}\cdot \gamma_{\mathrm{eff}}^2} \left[2\cdot q\cdot C\cdot \mathrm{acosh}\left( C \right)  \right.\right.\\
\left.\left. -2\cdot q\cdot \sqrt{(C^2 - 1)}\right]\right)
\label{SFArate}
\end{aligned}
\end{equation}
Here, the following definitions are used:
\begin{align}
A_{\mathrm{SFA}} &= E_{\mathrm{SFA}} / \omega_{\mathrm{SFA}} \\
q &= (A_{\mathrm{SFA}} + p_0) / A_{\mathrm{SFA}} \\
I_p^{\mathrm{eff}} &=I_p+0.5 \cdot p_{x0}^2 \\
\gamma_{\mathrm{eff}} &= \sqrt{(2 \cdot  I_p^{\mathrm{eff}})} / A_{\mathrm{SFA}} \\
C &= (1 + q^2 + \gamma_{\mathrm{eff}}^2) / (2 \cdot q) 
\end{align}
Therefore, the tunneling probability $P$ in our SCTS model incorporates non-adiabaticity \cite{Misha2005,Eckart2018_Offsets} and is a function of $E_{\mathrm{SFA}}$, $p_{x0}$ and $p_0$. The tunnel exit for each trajectory is obtained by solving Eq. (5) from Ref. \cite{Shilovski2012} and afterwards each semi-classical trajectory is propagated in the $1/r$-potential in the presence of the time-dependent laser electric field. We use a laser field with a total duration of 12 cycles that has a sine-square envelope and calculate 250 million semi-classical trajectories for the CoRTC and the CRTC scenario. Convergence of the simulations is tested by varying the bin size of the grid and the method of ``phase compression'' is used \cite{Eckart2019Diss,Eckart_TheoHASE2020}. The laser electric field in the SCTS model has been scaled down by 8\% to improve agreement of experiment and theory for low energy electrons.


\begin{thebibliography}{37}%
\makeatletter
\providecommand \@ifxundefined [1]{%
 \@ifx{#1\undefined}
}%
\providecommand \@ifnum [1]{%
 \ifnum #1\expandafter \@firstoftwo
 \else \expandafter \@secondoftwo
 \fi
}%
\providecommand \@ifx [1]{%
 \ifx #1\expandafter \@firstoftwo
 \else \expandafter \@secondoftwo
 \fi
}%
\providecommand \natexlab [1]{#1}%
\providecommand \enquote  [1]{``#1''}%
\providecommand \bibnamefont  [1]{#1}%
\providecommand \bibfnamefont [1]{#1}%
\providecommand \citenamefont [1]{#1}%
\providecommand \href@noop [0]{\@secondoftwo}%
\providecommand \href [0]{\begingroup \@sanitize@url \@href}%
\providecommand \@href[1]{\@@startlink{#1}\@@href}%
\providecommand \@@href[1]{\endgroup#1\@@endlink}%
\providecommand \@sanitize@url [0]{\catcode `\\12\catcode `\$12\catcode
  `\&12\catcode `\#12\catcode `\^12\catcode `\_12\catcode `\%12\relax}%
\providecommand \@@startlink[1]{}%
\providecommand \@@endlink[0]{}%
\providecommand \url  [0]{\begingroup\@sanitize@url \@url }%
\providecommand \@url [1]{\endgroup\@href {#1}{\urlprefix }}%
\providecommand \urlprefix  [0]{URL }%
\providecommand \Eprint [0]{\href }%
\providecommand \doibase [0]{http://dx.doi.org/}%
\providecommand \selectlanguage [0]{\@gobble}%
\providecommand \bibinfo  [0]{\@secondoftwo}%
\providecommand \bibfield  [0]{\@secondoftwo}%
\providecommand \translation [1]{[#1]}%
\providecommand \BibitemOpen [0]{}%
\providecommand \bibitemStop [0]{}%
\providecommand \bibitemNoStop [0]{.\EOS\space}%
\providecommand \EOS [0]{\spacefactor3000\relax}%
\providecommand \BibitemShut  [1]{\csname bibitem#1\endcsname}%
\let\auto@bib@innerbib\@empty
\bibitem [{\citenamefont {Keldysh}(1965)}]{Keldysh1965}%
  \BibitemOpen
  \bibfield  {author} {\bibinfo {author} {\bibfnamefont {L.~V.}\ \bibnamefont
  {Keldysh}},\ }\bibfield  {title} {\enquote {\bibinfo {title} {{Ionization in
  the field of a strong electromagnetic wave}},}\ }\href@noop {} {\bibfield
  {journal} {\bibinfo  {journal} {Sov. Phys. JETP}\ }\textbf {\bibinfo {volume}
  {20}},\ \bibinfo {pages} {1307} (\bibinfo {year} {1965})}\BibitemShut
  {NoStop}%
\bibitem [{\citenamefont {Voronov}\ and\ \citenamefont
  {Delone}(1966)}]{voronov1966many}%
  \BibitemOpen
  \bibfield  {author} {\bibinfo {author} {\bibfnamefont {G.~S.}\ \bibnamefont
  {Voronov}}\ and\ \bibinfo {author} {\bibfnamefont {N.~B.}\ \bibnamefont
  {Delone}},\ }\bibfield  {title} {\enquote {\bibinfo {title} {Many-photon
  ionization of the xenon atom by ruby laser radiation},}\ }\href@noop {}
  {\bibfield  {journal} {\bibinfo  {journal} {Sov. Phys. JETP}\ }\textbf
  {\bibinfo {volume} {23}},\ \bibinfo {pages} {54} (\bibinfo {year}
  {1966})}\BibitemShut {NoStop}%
\bibitem [{\citenamefont {Agostini}\ \emph {et~al.}(1979)\citenamefont
  {Agostini}, \citenamefont {Fabre}, \citenamefont {Mainfray}, \citenamefont
  {Petite},\ and\ \citenamefont {Rahman}}]{agostini1979p}%
  \BibitemOpen
  \bibfield  {author} {\bibinfo {author} {\bibfnamefont {P.}~\bibnamefont
  {Agostini}}, \bibinfo {author} {\bibfnamefont {F.}~\bibnamefont {Fabre}},
  \bibinfo {author} {\bibfnamefont {G.}~\bibnamefont {Mainfray}}, \bibinfo
  {author} {\bibfnamefont {G.}~\bibnamefont {Petite}}, \ and\ \bibinfo {author}
  {\bibfnamefont {N.~K.}\ \bibnamefont {Rahman}},\ }\bibfield  {title}
  {\enquote {\bibinfo {title} {Free-free transitions following six-photon
  ionization of xenon atoms},}\ }\href {\doibase 10.1103/PhysRevLett.42.1127}
  {\bibfield  {journal} {\bibinfo  {journal} {Phys. Rev. Lett.}\ }\textbf
  {\bibinfo {volume} {42}},\ \bibinfo {pages} {1127} (\bibinfo {year}
  {1979})}\BibitemShut {NoStop}%
\bibitem [{\citenamefont {Freeman}\ \emph {et~al.}(1987)\citenamefont
  {Freeman}, \citenamefont {Bucksbaum}, \citenamefont {Milchberg},
  \citenamefont {Darack}, \citenamefont {Schumacher},\ and\ \citenamefont
  {Geusic}}]{Freeman1987}%
  \BibitemOpen
  \bibfield  {author} {\bibinfo {author} {\bibfnamefont {R.~R.}\ \bibnamefont
  {Freeman}}, \bibinfo {author} {\bibfnamefont {P.~H.}\ \bibnamefont
  {Bucksbaum}}, \bibinfo {author} {\bibfnamefont {H.}~\bibnamefont
  {Milchberg}}, \bibinfo {author} {\bibfnamefont {S.}~\bibnamefont {Darack}},
  \bibinfo {author} {\bibfnamefont {D.}~\bibnamefont {Schumacher}}, \ and\
  \bibinfo {author} {\bibfnamefont {M.~E.}\ \bibnamefont {Geusic}},\ }\bibfield
   {title} {\enquote {\bibinfo {title} {Above-threshold ionization with
  subpicosecond laser pulses},}\ }\href {\doibase 10.1103/PhysRevLett.59.1092}
  {\bibfield  {journal} {\bibinfo  {journal} {Phys. Rev. Lett.}\ }\textbf
  {\bibinfo {volume} {59}},\ \bibinfo {pages} {1092} (\bibinfo {year}
  {1987})}\BibitemShut {NoStop}%
\bibitem [{\citenamefont {Petite}\ \emph {et~al.}(1988)\citenamefont {Petite},
  \citenamefont {Agostini},\ and\ \citenamefont {Muller}}]{Petite_ATI_1988}%
  \BibitemOpen
  \bibfield  {author} {\bibinfo {author} {\bibfnamefont {G.}~\bibnamefont
  {Petite}}, \bibinfo {author} {\bibfnamefont {P.}~\bibnamefont {Agostini}}, \
  and\ \bibinfo {author} {\bibfnamefont {H.G.}\ \bibnamefont {Muller}},\
  }\bibfield  {title} {\enquote {\bibinfo {title} {Intensity dependence of
  non-perturbative above-threshold ionisation spectra: experimental study},}\
  }\href {http://stacks.iop.org/0953-4075/21/i=24/a=010} {\bibfield  {journal}
  {\bibinfo  {journal} {J. Phys. B}\ }\textbf {\bibinfo {volume} {21}},\
  \bibinfo {pages} {4097} (\bibinfo {year} {1988})}\BibitemShut {NoStop}%
\bibitem [{\citenamefont {{{M. Yu. Ivanov}}}\ \emph {et~al.}(2005)\citenamefont
  {{{M. Yu. Ivanov}}}, \citenamefont {Spanner},\ and\ \citenamefont
  {Smirnova}}]{Misha2005}%
  \BibitemOpen
  \bibfield  {author} {\bibinfo {author} {\bibnamefont {{{M. Yu. Ivanov}}}},
  \bibinfo {author} {\bibfnamefont {M.}~\bibnamefont {Spanner}}, \ and\
  \bibinfo {author} {\bibfnamefont {O.}~\bibnamefont {Smirnova}},\ }\bibfield
  {title} {\enquote {\bibinfo {title} {Anatomy of strong field ionization},}\
  }\href {\doibase 10.1080/0950034042000275360} {\bibfield  {journal} {\bibinfo
   {journal} {J. Mod. Opt.}\ }\textbf {\bibinfo {volume} {52}},\ \bibinfo
  {pages} {165} (\bibinfo {year} {2005})}\BibitemShut {NoStop}%
\bibitem [{\citenamefont {Ge}\ \emph {et~al.}(2019)\citenamefont {Ge},
  \citenamefont {Han}, \citenamefont {Deng}, \citenamefont {Gong},\ and\
  \citenamefont {Liu}}]{Ge2019}%
  \BibitemOpen
  \bibfield  {author} {\bibinfo {author} {\bibfnamefont {P.}~\bibnamefont
  {Ge}}, \bibinfo {author} {\bibfnamefont {M.}~\bibnamefont {Han}}, \bibinfo
  {author} {\bibfnamefont {Y.}~\bibnamefont {Deng}}, \bibinfo {author}
  {\bibfnamefont {Q.}~\bibnamefont {Gong}}, \ and\ \bibinfo {author}
  {\bibfnamefont {Y.}~\bibnamefont {Liu}},\ }\bibfield  {title} {\enquote
  {\bibinfo {title} {Universal description of the attoclock with two-color
  corotating circular fields},}\ }\href {\doibase
  10.1103/PhysRevLett.122.013201} {\bibfield  {journal} {\bibinfo  {journal}
  {Phys. Rev. Lett.}\ }\textbf {\bibinfo {volume} {122}},\ \bibinfo {pages}
  {013201} (\bibinfo {year} {2019})}\BibitemShut {NoStop}%
\bibitem [{\citenamefont {Eichmann}\ \emph {et~al.}(1995)\citenamefont
  {Eichmann}, \citenamefont {Egbert}, \citenamefont {Nolte}, \citenamefont
  {Momma}, \citenamefont {Wellegehausen}, \citenamefont {Becker}, \citenamefont
  {Long},\ and\ \citenamefont {McIver}}]{Eichmann1995}%
  \BibitemOpen
  \bibfield  {author} {\bibinfo {author} {\bibfnamefont {H.}~\bibnamefont
  {Eichmann}}, \bibinfo {author} {\bibfnamefont {A.}~\bibnamefont {Egbert}},
  \bibinfo {author} {\bibfnamefont {S.}~\bibnamefont {Nolte}}, \bibinfo
  {author} {\bibfnamefont {C.}~\bibnamefont {Momma}}, \bibinfo {author}
  {\bibfnamefont {B.}~\bibnamefont {Wellegehausen}}, \bibinfo {author}
  {\bibfnamefont {W.}~\bibnamefont {Becker}}, \bibinfo {author} {\bibfnamefont
  {S.}~\bibnamefont {Long}}, \ and\ \bibinfo {author} {\bibfnamefont {J.~K.}\
  \bibnamefont {McIver}},\ }\bibfield  {title} {\enquote {\bibinfo {title}
  {Polarization-dependent high-order two-color mixing},}\ }\href {\doibase
  10.1103/PhysRevA.51.R3414} {\bibfield  {journal} {\bibinfo  {journal} {Phys.
  Rev. A}\ }\textbf {\bibinfo {volume} {51}},\ \bibinfo {pages} {R3414}
  (\bibinfo {year} {1995})}\BibitemShut {NoStop}%
\bibitem [{\citenamefont {Becker}\ \emph {et~al.}(1999)\citenamefont {Becker},
  \citenamefont {Chichkov},\ and\ \citenamefont {Wellegehausen}}]{Becker1999}%
  \BibitemOpen
  \bibfield  {author} {\bibinfo {author} {\bibfnamefont {W.}~\bibnamefont
  {Becker}}, \bibinfo {author} {\bibfnamefont {B.~N.}\ \bibnamefont
  {Chichkov}}, \ and\ \bibinfo {author} {\bibfnamefont {B.}~\bibnamefont
  {Wellegehausen}},\ }\bibfield  {title} {\enquote {\bibinfo {title} {Schemes
  for the generation of circularly polarized high-order harmonics by two-color
  mixing},}\ }\href {\doibase 10.1103/PhysRevA.60.1721} {\bibfield  {journal}
  {\bibinfo  {journal} {Phys. Rev. A}\ }\textbf {\bibinfo {volume} {60}},\
  \bibinfo {pages} {1721} (\bibinfo {year} {1999})}\BibitemShut {NoStop}%
\bibitem [{\citenamefont {Fleischer}\ \emph {et~al.}(2014)\citenamefont
  {Fleischer}, \citenamefont {Kfir}, \citenamefont {Diskin}, \citenamefont
  {Sidorenko},\ and\ \citenamefont {Cohen}}]{fleischer2014spin}%
  \BibitemOpen
  \bibfield  {author} {\bibinfo {author} {\bibfnamefont {A.}~\bibnamefont
  {Fleischer}}, \bibinfo {author} {\bibfnamefont {O.}~\bibnamefont {Kfir}},
  \bibinfo {author} {\bibfnamefont {T.}~\bibnamefont {Diskin}}, \bibinfo
  {author} {\bibfnamefont {P.}~\bibnamefont {Sidorenko}}, \ and\ \bibinfo
  {author} {\bibfnamefont {O.}~\bibnamefont {Cohen}},\ }\bibfield  {title}
  {\enquote {\bibinfo {title} {Spin angular momentum and tunable polarization
  in high-harmonic generation},}\ }\href@noop {} {\bibfield  {journal}
  {\bibinfo  {journal} {Nat. Photonics}\ }\textbf {\bibinfo {volume} {8}},\
  \bibinfo {pages} {543} (\bibinfo {year} {2014})}\BibitemShut {NoStop}%
\bibitem [{\citenamefont {Mancuso}\ \emph {et~al.}(2015)\citenamefont
  {Mancuso}, \citenamefont {Hickstein}, \citenamefont {Grychtol}, \citenamefont
  {Knut}, \citenamefont {Kfir}, \citenamefont {Tong}, \citenamefont {Dollar},
  \citenamefont {Zusin}, \citenamefont {Gopalakrishnan}, \citenamefont
  {Gentry}, \citenamefont {Turgut}, \citenamefont {Ellis}, \citenamefont
  {Chen}, \citenamefont {Fleischer}, \citenamefont {Cohen}, \citenamefont
  {Kapteyn},\ and\ \citenamefont {Murnane}}]{MancusoPRA2015_CRTC}%
  \BibitemOpen
  \bibfield  {author} {\bibinfo {author} {\bibfnamefont {C.~A.}\ \bibnamefont
  {Mancuso}}, \bibinfo {author} {\bibfnamefont {D.~D.}\ \bibnamefont
  {Hickstein}}, \bibinfo {author} {\bibfnamefont {P.}~\bibnamefont {Grychtol}},
  \bibinfo {author} {\bibfnamefont {R.}~\bibnamefont {Knut}}, \bibinfo {author}
  {\bibfnamefont {O.}~\bibnamefont {Kfir}}, \bibinfo {author} {\bibfnamefont
  {X.-M.}\ \bibnamefont {Tong}}, \bibinfo {author} {\bibfnamefont
  {F.}~\bibnamefont {Dollar}}, \bibinfo {author} {\bibfnamefont
  {D.}~\bibnamefont {Zusin}}, \bibinfo {author} {\bibfnamefont
  {M.}~\bibnamefont {Gopalakrishnan}}, \bibinfo {author} {\bibfnamefont
  {C.}~\bibnamefont {Gentry}}, \bibinfo {author} {\bibfnamefont
  {E.}~\bibnamefont {Turgut}}, \bibinfo {author} {\bibfnamefont {J.~L.}\
  \bibnamefont {Ellis}}, \bibinfo {author} {\bibfnamefont {M.-C.}\ \bibnamefont
  {Chen}}, \bibinfo {author} {\bibfnamefont {A.}~\bibnamefont {Fleischer}},
  \bibinfo {author} {\bibfnamefont {O.}~\bibnamefont {Cohen}}, \bibinfo
  {author} {\bibfnamefont {H.~C.}\ \bibnamefont {Kapteyn}}, \ and\ \bibinfo
  {author} {\bibfnamefont {M.~M.}\ \bibnamefont {Murnane}},\ }\bibfield
  {title} {\enquote {\bibinfo {title} {Strong-field ionization with two-color
  circularly polarized laser fields},}\ }\href {\doibase
  10.1103/PhysRevA.91.031402} {\bibfield  {journal} {\bibinfo  {journal} {Phys.
  Rev. A}\ }\textbf {\bibinfo {volume} {91}},\ \bibinfo {pages} {031402(R)}
  (\bibinfo {year} {2015})}\BibitemShut {NoStop}%
\bibitem [{\citenamefont {Han}\ \emph {et~al.}(2018)\citenamefont {Han},
  \citenamefont {Ge}, \citenamefont {Shao}, \citenamefont {Gong},\ and\
  \citenamefont {Liu}}]{Han2018}%
  \BibitemOpen
  \bibfield  {author} {\bibinfo {author} {\bibfnamefont {M.}~\bibnamefont
  {Han}}, \bibinfo {author} {\bibfnamefont {P.}~\bibnamefont {Ge}}, \bibinfo
  {author} {\bibfnamefont {Y.}~\bibnamefont {Shao}}, \bibinfo {author}
  {\bibfnamefont {Q.}~\bibnamefont {Gong}}, \ and\ \bibinfo {author}
  {\bibfnamefont {Y.}~\bibnamefont {Liu}},\ }\bibfield  {title} {\enquote
  {\bibinfo {title} {Attoclock photoelectron interferometry with two-color
  corotating circular fields to probe the phase and the amplitude of emitting
  wave packets},}\ }\href {\doibase 10.1103/PhysRevLett.120.073202} {\bibfield
  {journal} {\bibinfo  {journal} {Phys. Rev. Lett.}\ }\textbf {\bibinfo
  {volume} {120}},\ \bibinfo {pages} {073202} (\bibinfo {year}
  {2018})}\BibitemShut {NoStop}%
\bibitem [{\citenamefont {Eckart}(2020)}]{Eckart_TheoHASE2020}%
  \BibitemOpen
  \bibfield  {author} {\bibinfo {author} {\bibfnamefont {S.}~\bibnamefont
  {Eckart}},\ }\bibfield  {title} {\enquote {\bibinfo {title} {Holographic
  angular streaking of electrons and the {{Wigner}} time delay},}\ }\href
  {\doibase 10.1103/PhysRevResearch.2.033248} {\bibfield  {journal} {\bibinfo
  {journal} {Phys. Rev. Research}\ }\textbf {\bibinfo {volume} {2}},\ \bibinfo
  {pages} {033248} (\bibinfo {year} {2020})}\BibitemShut {NoStop}%
\bibitem [{\citenamefont {Trabert}\ \emph {et~al.}(2020)\citenamefont
  {Trabert}, \citenamefont {Fehre}, \citenamefont {Anders}, \citenamefont
  {Geyer}, \citenamefont {Grundmann}, \citenamefont {Sch\"offler},
  \citenamefont {{{L. Ph. H. Schmidt}}}, \citenamefont {Jahnke}, \citenamefont
  {D\"orner}, \citenamefont {Kunitski},\ and\ \citenamefont
  {Eckart}}]{DanielArXiv2020}%
  \BibitemOpen
  \bibfield  {author} {\bibinfo {author} {\bibfnamefont {D.}~\bibnamefont
  {Trabert}}, \bibinfo {author} {\bibfnamefont {K.}~\bibnamefont {Fehre}},
  \bibinfo {author} {\bibfnamefont {N.}~\bibnamefont {Anders}}, \bibinfo
  {author} {\bibfnamefont {A.}~\bibnamefont {Geyer}}, \bibinfo {author}
  {\bibfnamefont {S.}~\bibnamefont {Grundmann}}, \bibinfo {author}
  {\bibfnamefont {M.}~\bibnamefont {Sch\"offler}}, \bibinfo {author}
  {\bibnamefont {{{L. Ph. H. Schmidt}}}}, \bibinfo {author} {\bibfnamefont
  {T.}~\bibnamefont {Jahnke}}, \bibinfo {author} {\bibfnamefont
  {R.}~\bibnamefont {D\"orner}}, \bibinfo {author} {\bibfnamefont
  {M.}~\bibnamefont {Kunitski}}, \ and\ \bibinfo {author} {\bibfnamefont
  {S.}~\bibnamefont {Eckart}},\ }\bibfield  {title} {\enquote {\bibinfo {title}
  {Angular dependence of the {{Wigner}} time delay upon tunnel ionization of
  {{H}}$_2$},}\ }\href@noop {} {\bibfield  {journal} {\bibinfo  {journal}
  {arXiv preprint, arXiv:2005.09584}\ } (\bibinfo {year} {2020})}\BibitemShut
  {NoStop}%
\bibitem [{\citenamefont {Eckart}\ \emph
  {et~al.}(2018{\natexlab{a}})\citenamefont {Eckart}, \citenamefont {Fehre},
  \citenamefont {Eicke}, \citenamefont {Hartung}, \citenamefont {Rist},
  \citenamefont {Trabert}, \citenamefont {Strenger}, \citenamefont {Pier},
  \citenamefont {{{L. Ph. H. Schmidt}}}, \citenamefont {Jahnke}, \citenamefont
  {Sch\"offler}, \citenamefont {Lein}, \citenamefont {Kunitski},\ and\
  \citenamefont {D\"orner}}]{Eckart2018_Offsets}%
  \BibitemOpen
  \bibfield  {author} {\bibinfo {author} {\bibfnamefont {S.}~\bibnamefont
  {Eckart}}, \bibinfo {author} {\bibfnamefont {K.}~\bibnamefont {Fehre}},
  \bibinfo {author} {\bibfnamefont {N.}~\bibnamefont {Eicke}}, \bibinfo
  {author} {\bibfnamefont {A.}~\bibnamefont {Hartung}}, \bibinfo {author}
  {\bibfnamefont {J.}~\bibnamefont {Rist}}, \bibinfo {author} {\bibfnamefont
  {D.}~\bibnamefont {Trabert}}, \bibinfo {author} {\bibfnamefont
  {N.}~\bibnamefont {Strenger}}, \bibinfo {author} {\bibfnamefont
  {A.}~\bibnamefont {Pier}}, \bibinfo {author} {\bibnamefont {{{L. Ph. H.
  Schmidt}}}}, \bibinfo {author} {\bibfnamefont {T.}~\bibnamefont {Jahnke}},
  \bibinfo {author} {\bibfnamefont {M.~S.}\ \bibnamefont {Sch\"offler}},
  \bibinfo {author} {\bibfnamefont {M.}~\bibnamefont {Lein}}, \bibinfo {author}
  {\bibfnamefont {M.}~\bibnamefont {Kunitski}}, \ and\ \bibinfo {author}
  {\bibfnamefont {R.}~\bibnamefont {D\"orner}},\ }\bibfield  {title} {\enquote
  {\bibinfo {title} {Direct experimental access to the nonadiabatic initial
  momentum offset upon tunnel ionization},}\ }\href {\doibase
  10.1103/PhysRevLett.121.163202} {\bibfield  {journal} {\bibinfo  {journal}
  {Phys. Rev. Lett.}\ }\textbf {\bibinfo {volume} {121}},\ \bibinfo {pages}
  {163202} (\bibinfo {year} {2018}{\natexlab{a}})}\BibitemShut {NoStop}%
\bibitem [{\citenamefont {Zhang}\ \emph {et~al.}(2014)\citenamefont {Zhang},
  \citenamefont {Xie}, \citenamefont {Roither}, \citenamefont {Kartashov},
  \citenamefont {Wang}, \citenamefont {Wang}, \citenamefont {Sch\"offler},
  \citenamefont {Shafir}, \citenamefont {Corkum}, \citenamefont
  {Baltu\ifmmode\check{s}\else\v{s}\fi{}ka}, \citenamefont {Ivanov},
  \citenamefont {Kheifets}, \citenamefont {Liu}, \citenamefont {Staudte},\ and\
  \citenamefont {Kitzler}}]{Zhang2014}%
  \BibitemOpen
  \bibfield  {author} {\bibinfo {author} {\bibfnamefont {L.}~\bibnamefont
  {Zhang}}, \bibinfo {author} {\bibfnamefont {X.}~\bibnamefont {Xie}}, \bibinfo
  {author} {\bibfnamefont {S.}~\bibnamefont {Roither}}, \bibinfo {author}
  {\bibfnamefont {D.}~\bibnamefont {Kartashov}}, \bibinfo {author}
  {\bibfnamefont {Y.~L.}\ \bibnamefont {Wang}}, \bibinfo {author}
  {\bibfnamefont {C.~L.}\ \bibnamefont {Wang}}, \bibinfo {author}
  {\bibfnamefont {M.}~\bibnamefont {Sch\"offler}}, \bibinfo {author}
  {\bibfnamefont {D.}~\bibnamefont {Shafir}}, \bibinfo {author} {\bibfnamefont
  {P.~B.}\ \bibnamefont {Corkum}}, \bibinfo {author} {\bibfnamefont
  {A.}~\bibnamefont {Baltu\ifmmode\check{s}\else\v{s}\fi{}ka}}, \bibinfo
  {author} {\bibfnamefont {I.}~\bibnamefont {Ivanov}}, \bibinfo {author}
  {\bibfnamefont {A.}~\bibnamefont {Kheifets}}, \bibinfo {author}
  {\bibfnamefont {X.~J.}\ \bibnamefont {Liu}}, \bibinfo {author} {\bibfnamefont
  {A.}~\bibnamefont {Staudte}}, \ and\ \bibinfo {author} {\bibfnamefont
  {M.}~\bibnamefont {Kitzler}},\ }\bibfield  {title} {\enquote {\bibinfo
  {title} {Laser-sub-cycle two-dimensional electron-momentum mapping using
  orthogonal two-color fields},}\ }\href {\doibase 10.1103/PhysRevA.90.061401}
  {\bibfield  {journal} {\bibinfo  {journal} {Phys. Rev. A}\ }\textbf {\bibinfo
  {volume} {90}},\ \bibinfo {pages} {061401(R)} (\bibinfo {year}
  {2014})}\BibitemShut {NoStop}%
\bibitem [{\citenamefont {Richter}\ \emph {et~al.}(2015)\citenamefont
  {Richter}, \citenamefont {Kunitski}, \citenamefont {Sch\"offler},
  \citenamefont {Jahnke}, \citenamefont {{{L. Ph. H. Schmidt}}}, \citenamefont
  {Li}, \citenamefont {Liu},\ and\ \citenamefont {D\"orner}}]{Richter2015}%
  \BibitemOpen
  \bibfield  {author} {\bibinfo {author} {\bibfnamefont {M.}~\bibnamefont
  {Richter}}, \bibinfo {author} {\bibfnamefont {M.}~\bibnamefont {Kunitski}},
  \bibinfo {author} {\bibfnamefont {M.}~\bibnamefont {Sch\"offler}}, \bibinfo
  {author} {\bibfnamefont {T.}~\bibnamefont {Jahnke}}, \bibinfo {author}
  {\bibnamefont {{{L. Ph. H. Schmidt}}}}, \bibinfo {author} {\bibfnamefont
  {M.}~\bibnamefont {Li}}, \bibinfo {author} {\bibfnamefont {Y.}~\bibnamefont
  {Liu}}, \ and\ \bibinfo {author} {\bibfnamefont {R.}~\bibnamefont
  {D\"orner}},\ }\bibfield  {title} {\enquote {\bibinfo {title} {Streaking
  temporal double-slit interference by an orthogonal two-color laser field},}\
  }\href {\doibase 10.1103/PhysRevLett.114.143001} {\bibfield  {journal}
  {\bibinfo  {journal} {Phys. Rev. Lett.}\ }\textbf {\bibinfo {volume} {114}},\
  \bibinfo {pages} {143001} (\bibinfo {year} {2015})}\BibitemShut {NoStop}%
\bibitem [{\citenamefont {Eckart}\ \emph
  {et~al.}(2018{\natexlab{b}})\citenamefont {Eckart}, \citenamefont {Kunitski},
  \citenamefont {Ivanov}, \citenamefont {Richter}, \citenamefont {Fehre},
  \citenamefont {Hartung}, \citenamefont {Rist}, \citenamefont {Henrichs},
  \citenamefont {Trabert}, \citenamefont {Schlott}, \citenamefont {{{L. Ph. H.
  Schmidt}}}, \citenamefont {Jahnke}, \citenamefont {Sch\"offler},
  \citenamefont {Kheifets},\ and\ \citenamefont
  {D\"orner}}]{Eckart2018SubCycle}%
  \BibitemOpen
  \bibfield  {author} {\bibinfo {author} {\bibfnamefont {S.}~\bibnamefont
  {Eckart}}, \bibinfo {author} {\bibfnamefont {M.}~\bibnamefont {Kunitski}},
  \bibinfo {author} {\bibfnamefont {I.}~\bibnamefont {Ivanov}}, \bibinfo
  {author} {\bibfnamefont {M.}~\bibnamefont {Richter}}, \bibinfo {author}
  {\bibfnamefont {K.}~\bibnamefont {Fehre}}, \bibinfo {author} {\bibfnamefont
  {A.}~\bibnamefont {Hartung}}, \bibinfo {author} {\bibfnamefont
  {J.}~\bibnamefont {Rist}}, \bibinfo {author} {\bibfnamefont {K.}~\bibnamefont
  {Henrichs}}, \bibinfo {author} {\bibfnamefont {D.}~\bibnamefont {Trabert}},
  \bibinfo {author} {\bibfnamefont {N.}~\bibnamefont {Schlott}}, \bibinfo
  {author} {\bibnamefont {{{L. Ph. H. Schmidt}}}}, \bibinfo {author}
  {\bibfnamefont {T.}~\bibnamefont {Jahnke}}, \bibinfo {author} {\bibfnamefont
  {M.~S.}\ \bibnamefont {Sch\"offler}}, \bibinfo {author} {\bibfnamefont
  {A.}~\bibnamefont {Kheifets}}, \ and\ \bibinfo {author} {\bibfnamefont
  {R.}~\bibnamefont {D\"orner}},\ }\bibfield  {title} {\enquote {\bibinfo
  {title} {Subcycle interference upon tunnel ionization by counter-rotating
  two-color fields},}\ }\href {\doibase 10.1103/PhysRevA.97.041402} {\bibfield
  {journal} {\bibinfo  {journal} {Phys. Rev. A}\ }\textbf {\bibinfo {volume}
  {97}},\ \bibinfo {pages} {041402(R)} (\bibinfo {year}
  {2018}{\natexlab{b}})}\BibitemShut {NoStop}%
\bibitem [{\citenamefont {Mancuso}\ \emph {et~al.}(2016)\citenamefont
  {Mancuso}, \citenamefont {Dorney}, \citenamefont {Hickstein}, \citenamefont
  {Chaloupka}, \citenamefont {Ellis}, \citenamefont {Dollar}, \citenamefont
  {Knut}, \citenamefont {Grychtol}, \citenamefont {Zusin}, \citenamefont
  {Gentry}, \citenamefont {Gopalakrishnan}, \citenamefont {Kapteyn},\ and\
  \citenamefont {Murnane}}]{Mancuso2016PRL}%
  \BibitemOpen
  \bibfield  {author} {\bibinfo {author} {\bibfnamefont {C.~A.}\ \bibnamefont
  {Mancuso}}, \bibinfo {author} {\bibfnamefont {K.~M.}\ \bibnamefont {Dorney}},
  \bibinfo {author} {\bibfnamefont {D.~D.}\ \bibnamefont {Hickstein}}, \bibinfo
  {author} {\bibfnamefont {J.~L.}\ \bibnamefont {Chaloupka}}, \bibinfo {author}
  {\bibfnamefont {J.~L.}\ \bibnamefont {Ellis}}, \bibinfo {author}
  {\bibfnamefont {F.~J.}\ \bibnamefont {Dollar}}, \bibinfo {author}
  {\bibfnamefont {R.}~\bibnamefont {Knut}}, \bibinfo {author} {\bibfnamefont
  {P.}~\bibnamefont {Grychtol}}, \bibinfo {author} {\bibfnamefont
  {D.}~\bibnamefont {Zusin}}, \bibinfo {author} {\bibfnamefont
  {C.}~\bibnamefont {Gentry}}, \bibinfo {author} {\bibfnamefont
  {M.}~\bibnamefont {Gopalakrishnan}}, \bibinfo {author} {\bibfnamefont
  {H.~C.}\ \bibnamefont {Kapteyn}}, \ and\ \bibinfo {author} {\bibfnamefont
  {M.~M.}\ \bibnamefont {Murnane}},\ }\bibfield  {title} {\enquote {\bibinfo
  {title} {Controlling nonsequential double ionization in two-color circularly
  polarized femtosecond laser fields},}\ }\href {\doibase
  10.1103/PhysRevLett.117.133201} {\bibfield  {journal} {\bibinfo  {journal}
  {Phys. Rev. Lett.}\ }\textbf {\bibinfo {volume} {117}},\ \bibinfo {pages}
  {133201} (\bibinfo {year} {2016})}\BibitemShut {NoStop}%
\bibitem [{\citenamefont {Eckart}\ \emph {et~al.}(2016)\citenamefont {Eckart},
  \citenamefont {Richter}, \citenamefont {Kunitski}, \citenamefont {Hartung},
  \citenamefont {Rist}, \citenamefont {Henrichs}, \citenamefont {Schlott},
  \citenamefont {Kang}, \citenamefont {Bauer}, \citenamefont {Sann},
  \citenamefont {{{L. Ph. H. Schmidt}}}, \citenamefont {Sch\"offler},
  \citenamefont {Jahnke},\ and\ \citenamefont {D\"orner}}]{Eckart2016}%
  \BibitemOpen
  \bibfield  {author} {\bibinfo {author} {\bibfnamefont {S.}~\bibnamefont
  {Eckart}}, \bibinfo {author} {\bibfnamefont {M.}~\bibnamefont {Richter}},
  \bibinfo {author} {\bibfnamefont {M.}~\bibnamefont {Kunitski}}, \bibinfo
  {author} {\bibfnamefont {A.}~\bibnamefont {Hartung}}, \bibinfo {author}
  {\bibfnamefont {J.}~\bibnamefont {Rist}}, \bibinfo {author} {\bibfnamefont
  {K.}~\bibnamefont {Henrichs}}, \bibinfo {author} {\bibfnamefont
  {N.}~\bibnamefont {Schlott}}, \bibinfo {author} {\bibfnamefont
  {H.}~\bibnamefont {Kang}}, \bibinfo {author} {\bibfnamefont {T.}~\bibnamefont
  {Bauer}}, \bibinfo {author} {\bibfnamefont {H.}~\bibnamefont {Sann}},
  \bibinfo {author} {\bibnamefont {{{L. Ph. H. Schmidt}}}}, \bibinfo {author}
  {\bibfnamefont {M.}~\bibnamefont {Sch\"offler}}, \bibinfo {author}
  {\bibfnamefont {T.}~\bibnamefont {Jahnke}}, \ and\ \bibinfo {author}
  {\bibfnamefont {R.}~\bibnamefont {D\"orner}},\ }\bibfield  {title} {\enquote
  {\bibinfo {title} {Nonsequential double ionization by counterrotating
  circularly polarized two-color laser fields},}\ }\href {\doibase
  10.1103/PhysRevLett.117.133202} {\bibfield  {journal} {\bibinfo  {journal}
  {Phys. Rev. Lett.}\ }\textbf {\bibinfo {volume} {117}},\ \bibinfo {pages}
  {133202} (\bibinfo {year} {2016})}\BibitemShut {NoStop}%
\bibitem [{\citenamefont {Lin}\ \emph {et~al.}(2017)\citenamefont {Lin},
  \citenamefont {Jia}, \citenamefont {Yu}, \citenamefont {He}, \citenamefont
  {Ma}, \citenamefont {Li}, \citenamefont {Gong}, \citenamefont {Song},
  \citenamefont {Ji}, \citenamefont {Zhang}, \citenamefont {Li}, \citenamefont
  {Lu}, \citenamefont {Zeng}, \citenamefont {Chen},\ and\ \citenamefont
  {Wu}}]{Kang2017}%
  \BibitemOpen
  \bibfield  {author} {\bibinfo {author} {\bibfnamefont {K.}~\bibnamefont
  {Lin}}, \bibinfo {author} {\bibfnamefont {X.}~\bibnamefont {Jia}}, \bibinfo
  {author} {\bibfnamefont {Z.}~\bibnamefont {Yu}}, \bibinfo {author}
  {\bibfnamefont {F.}~\bibnamefont {He}}, \bibinfo {author} {\bibfnamefont
  {J.}~\bibnamefont {Ma}}, \bibinfo {author} {\bibfnamefont {H.}~\bibnamefont
  {Li}}, \bibinfo {author} {\bibfnamefont {X.}~\bibnamefont {Gong}}, \bibinfo
  {author} {\bibfnamefont {Q.}~\bibnamefont {Song}}, \bibinfo {author}
  {\bibfnamefont {Q.}~\bibnamefont {Ji}}, \bibinfo {author} {\bibfnamefont
  {W.}~\bibnamefont {Zhang}}, \bibinfo {author} {\bibfnamefont
  {H.}~\bibnamefont {Li}}, \bibinfo {author} {\bibfnamefont {P.}~\bibnamefont
  {Lu}}, \bibinfo {author} {\bibfnamefont {H.}~\bibnamefont {Zeng}}, \bibinfo
  {author} {\bibfnamefont {J.}~\bibnamefont {Chen}}, \ and\ \bibinfo {author}
  {\bibfnamefont {J.}~\bibnamefont {Wu}},\ }\bibfield  {title} {\enquote
  {\bibinfo {title} {Comparison study of strong-field ionization of molecules
  and atoms by bicircular two-color femtosecond laser pulses},}\ }\href
  {\doibase 10.1103/PhysRevLett.119.203202} {\bibfield  {journal} {\bibinfo
  {journal} {Phys. Rev. Lett.}\ }\textbf {\bibinfo {volume} {119}},\ \bibinfo
  {pages} {203202} (\bibinfo {year} {2017})}\BibitemShut {NoStop}%
\bibitem [{\citenamefont {Kerbstadt}\ \emph {et~al.}(2017)\citenamefont
  {Kerbstadt}, \citenamefont {Pengel}, \citenamefont {Johannmeyer},
  \citenamefont {Englert}, \citenamefont {Bayer},\ and\ \citenamefont
  {Wollenhaupt}}]{Kerbstadt2017}%
  \BibitemOpen
  \bibfield  {author} {\bibinfo {author} {\bibfnamefont {S.}~\bibnamefont
  {Kerbstadt}}, \bibinfo {author} {\bibfnamefont {D.}~\bibnamefont {Pengel}},
  \bibinfo {author} {\bibfnamefont {D.}~\bibnamefont {Johannmeyer}}, \bibinfo
  {author} {\bibfnamefont {L.}~\bibnamefont {Englert}}, \bibinfo {author}
  {\bibfnamefont {T.}~\bibnamefont {Bayer}}, \ and\ \bibinfo {author}
  {\bibfnamefont {M.}~\bibnamefont {Wollenhaupt}},\ }\bibfield  {title}
  {\enquote {\bibinfo {title} {Control of photoelectron momentum distributions
  by bichromatic polarization-shaped laser fields},}\ }\href {\doibase
  10.1088/1367-2630/aa83a4} {\bibfield  {journal} {\bibinfo  {journal} {New J.
  Phys.}\ }\textbf {\bibinfo {volume} {19}},\ \bibinfo {pages} {103017}
  (\bibinfo {year} {2017})}\BibitemShut {NoStop}%
\bibitem [{\citenamefont {Mancuso}\ \emph {et~al.}(2017)\citenamefont
  {Mancuso}, \citenamefont {Dorney}, \citenamefont {Hickstein}, \citenamefont
  {Chaloupka}, \citenamefont {Tong}, \citenamefont {Ellis}, \citenamefont
  {Kapteyn},\ and\ \citenamefont {Murnane}}]{Mancuso2017_enhancement}%
  \BibitemOpen
  \bibfield  {author} {\bibinfo {author} {\bibfnamefont {C.~A.}\ \bibnamefont
  {Mancuso}}, \bibinfo {author} {\bibfnamefont {K.~M.}\ \bibnamefont {Dorney}},
  \bibinfo {author} {\bibfnamefont {D.~D.}\ \bibnamefont {Hickstein}}, \bibinfo
  {author} {\bibfnamefont {J.~L.}\ \bibnamefont {Chaloupka}}, \bibinfo {author}
  {\bibfnamefont {X.-M.}\ \bibnamefont {Tong}}, \bibinfo {author}
  {\bibfnamefont {J.~L.}\ \bibnamefont {Ellis}}, \bibinfo {author}
  {\bibfnamefont {H.~C.}\ \bibnamefont {Kapteyn}}, \ and\ \bibinfo {author}
  {\bibfnamefont {M.~M.}\ \bibnamefont {Murnane}},\ }\bibfield  {title}
  {\enquote {\bibinfo {title} {Observation of ionization enhancement in
  two-color circularly polarized laser fields},}\ }\href {\doibase
  10.1103/PhysRevA.96.023402} {\bibfield  {journal} {\bibinfo  {journal} {Phys.
  Rev. A}\ }\textbf {\bibinfo {volume} {96}},\ \bibinfo {pages} {023402}
  (\bibinfo {year} {2017})}\BibitemShut {NoStop}%
\bibitem [{\citenamefont {Eckart}\ \emph
  {et~al.}(2018{\natexlab{c}})\citenamefont {Eckart}, \citenamefont {Kunitski},
  \citenamefont {Richter}, \citenamefont {Hartung}, \citenamefont {Rist},
  \citenamefont {Trinter}, \citenamefont {Fehre}, \citenamefont {Schlott},
  \citenamefont {Henrichs}, \citenamefont {{{L. Ph. H. Schmidt}}},
  \citenamefont {Jahnke}, \citenamefont {Sch{\"{o}}ffler}, \citenamefont {Liu},
  \citenamefont {Barth}, \citenamefont {Kaushal}, \citenamefont {Morales},
  \citenamefont {Ivanov}, \citenamefont {Smirnova},\ and\ \citenamefont
  {D{\"{o}}rner}}]{EckartNatPhys2018}%
  \BibitemOpen
  \bibfield  {author} {\bibinfo {author} {\bibfnamefont {S.}~\bibnamefont
  {Eckart}}, \bibinfo {author} {\bibfnamefont {M.}~\bibnamefont {Kunitski}},
  \bibinfo {author} {\bibfnamefont {M.}~\bibnamefont {Richter}}, \bibinfo
  {author} {\bibfnamefont {A.}~\bibnamefont {Hartung}}, \bibinfo {author}
  {\bibfnamefont {J.}~\bibnamefont {Rist}}, \bibinfo {author} {\bibfnamefont
  {F.}~\bibnamefont {Trinter}}, \bibinfo {author} {\bibfnamefont
  {K.}~\bibnamefont {Fehre}}, \bibinfo {author} {\bibfnamefont
  {N.}~\bibnamefont {Schlott}}, \bibinfo {author} {\bibfnamefont
  {K.}~\bibnamefont {Henrichs}}, \bibinfo {author} {\bibnamefont {{{L. Ph. H.
  Schmidt}}}}, \bibinfo {author} {\bibfnamefont {T.}~\bibnamefont {Jahnke}},
  \bibinfo {author} {\bibfnamefont {M.}~\bibnamefont {Sch{\"{o}}ffler}},
  \bibinfo {author} {\bibfnamefont {K.}~\bibnamefont {Liu}}, \bibinfo {author}
  {\bibfnamefont {I.}~\bibnamefont {Barth}}, \bibinfo {author} {\bibfnamefont
  {J.}~\bibnamefont {Kaushal}}, \bibinfo {author} {\bibfnamefont
  {F.}~\bibnamefont {Morales}}, \bibinfo {author} {\bibfnamefont
  {M.}~\bibnamefont {Ivanov}}, \bibinfo {author} {\bibfnamefont
  {O.}~\bibnamefont {Smirnova}}, \ and\ \bibinfo {author} {\bibfnamefont
  {R.}~\bibnamefont {D{\"{o}}rner}},\ }\bibfield  {title} {\enquote {\bibinfo
  {title} {{Ultrafast preparation and detection of ring currents in single
  atoms}},}\ }\href {\doibase 10.1038/s41567-018-0080-5} {\bibfield  {journal}
  {\bibinfo  {journal} {Nat. Phys.}\ }\textbf {\bibinfo {volume} {14}},\
  \bibinfo {pages} {701} (\bibinfo {year} {2018}{\natexlab{c}})}\BibitemShut
  {NoStop}%
\bibitem [{\citenamefont {Ullrich}\ \emph {et~al.}(2003)\citenamefont
  {Ullrich}, \citenamefont {Moshammer}, \citenamefont {Dorn}, \citenamefont
  {D{\"o}rner}, \citenamefont {{{L. Ph. H. Schmidt}}},\ and\ \citenamefont
  {Schmidt-B{\"o}cking}}]{ullrich2003recoil}%
  \BibitemOpen
  \bibfield  {author} {\bibinfo {author} {\bibfnamefont {J.}~\bibnamefont
  {Ullrich}}, \bibinfo {author} {\bibfnamefont {R.}~\bibnamefont {Moshammer}},
  \bibinfo {author} {\bibfnamefont {A.}~\bibnamefont {Dorn}}, \bibinfo {author}
  {\bibfnamefont {R.}~\bibnamefont {D{\"o}rner}}, \bibinfo {author}
  {\bibnamefont {{{L. Ph. H. Schmidt}}}}, \ and\ \bibinfo {author}
  {\bibfnamefont {H.}~\bibnamefont {Schmidt-B{\"o}cking}},\ }\bibfield  {title}
  {\enquote {\bibinfo {title} {Recoil-ion and electron momentum spectroscopy:
  reaction-microscopes},}\ }\href@noop {} {\bibfield  {journal} {\bibinfo
  {journal} {Rep. Prog. Phys.}\ }\textbf {\bibinfo {volume} {66}},\ \bibinfo
  {pages} {1463} (\bibinfo {year} {2003})}\BibitemShut {NoStop}%
\bibitem [{\citenamefont {Jagutzki}\ \emph {et~al.}(2002)\citenamefont
  {Jagutzki}, \citenamefont {Cerezo}, \citenamefont {Czasch}, \citenamefont
  {D{\"o}rner}, \citenamefont {Hattas}, \citenamefont {Huang}, \citenamefont
  {Mergel}, \citenamefont {Spillmann}, \citenamefont {Ullmann-Pfleger},
  \citenamefont {Weber}, \citenamefont {Schmidt-B{\"o}cking},\ and\
  \citenamefont {Smith}}]{jagutzki2002multiple}%
  \BibitemOpen
  \bibfield  {author} {\bibinfo {author} {\bibfnamefont {O.}~\bibnamefont
  {Jagutzki}}, \bibinfo {author} {\bibfnamefont {A.}~\bibnamefont {Cerezo}},
  \bibinfo {author} {\bibfnamefont {A.}~\bibnamefont {Czasch}}, \bibinfo
  {author} {\bibfnamefont {R.}~\bibnamefont {D{\"o}rner}}, \bibinfo {author}
  {\bibfnamefont {M.}~\bibnamefont {Hattas}}, \bibinfo {author} {\bibfnamefont
  {M.}~\bibnamefont {Huang}}, \bibinfo {author} {\bibfnamefont
  {V.}~\bibnamefont {Mergel}}, \bibinfo {author} {\bibfnamefont
  {U.}~\bibnamefont {Spillmann}}, \bibinfo {author} {\bibfnamefont
  {K.}~\bibnamefont {Ullmann-Pfleger}}, \bibinfo {author} {\bibfnamefont
  {T.}~\bibnamefont {Weber}}, \bibinfo {author} {\bibfnamefont
  {H.}~\bibnamefont {Schmidt-B{\"o}cking}}, \ and\ \bibinfo {author}
  {\bibfnamefont {G.~D.~W.}\ \bibnamefont {Smith}},\ }\bibfield  {title}
  {\enquote {\bibinfo {title} {Multiple hit readout of a microchannel plate
  detector with a three-layer delay-line anode},}\ }\href@noop {} {\bibfield
  {journal} {\bibinfo  {journal} {IEEE Trans. Nucl. Sci.}\ }\textbf {\bibinfo
  {volume} {49}},\ \bibinfo {pages} {2477} (\bibinfo {year}
  {2002})}\BibitemShut {NoStop}%
\bibitem [{\citenamefont {Barth}\ and\ \citenamefont
  {Smirnova}(2011)}]{Olga2011A}%
  \BibitemOpen
  \bibfield  {author} {\bibinfo {author} {\bibfnamefont {I.}~\bibnamefont
  {Barth}}\ and\ \bibinfo {author} {\bibfnamefont {O.}~\bibnamefont
  {Smirnova}},\ }\bibfield  {title} {\enquote {\bibinfo {title} {Nonadiabatic
  tunneling in circularly polarized laser fields: {Physical} picture and
  calculations},}\ }\href {\doibase 10.1103/PhysRevA.84.063415} {\bibfield
  {journal} {\bibinfo  {journal} {Phys. Rev. A}\ }\textbf {\bibinfo {volume}
  {84}},\ \bibinfo {pages} {063415} (\bibinfo {year} {2011})}\BibitemShut
  {NoStop}%
\bibitem [{\citenamefont {Barth}\ and\ \citenamefont
  {Smirnova}(2013)}]{Olga2011B}%
  \BibitemOpen
  \bibfield  {author} {\bibinfo {author} {\bibfnamefont {I.}~\bibnamefont
  {Barth}}\ and\ \bibinfo {author} {\bibfnamefont {O.}~\bibnamefont
  {Smirnova}},\ }\bibfield  {title} {\enquote {\bibinfo {title} {Nonadiabatic
  tunneling in circularly polarized laser fields. {II. Derivation} of
  formulas},}\ }\href {\doibase 10.1103/PhysRevA.87.013433} {\bibfield
  {journal} {\bibinfo  {journal} {Phys. Rev. A}\ }\textbf {\bibinfo {volume}
  {87}},\ \bibinfo {pages} {013433} (\bibinfo {year} {2013})}\BibitemShut
  {NoStop}%
\bibitem [{\citenamefont {Fano}(1985)}]{Fano1985}%
  \BibitemOpen
  \bibfield  {author} {\bibinfo {author} {\bibfnamefont {U.}~\bibnamefont
  {Fano}},\ }\bibfield  {title} {\enquote {\bibinfo {title} {Propensity rules:
  An analytical approach},}\ }\href {\doibase 10.1103/PhysRevA.32.617}
  {\bibfield  {journal} {\bibinfo  {journal} {Phys. Rev. A}\ }\textbf {\bibinfo
  {volume} {32}},\ \bibinfo {pages} {617--618} (\bibinfo {year}
  {1985})}\BibitemShut {NoStop}%
\bibitem [{\citenamefont {Arb\'o}\ \emph {et~al.}(2010)\citenamefont {Arb\'o},
  \citenamefont {Ishikawa}, \citenamefont {Schiessl}, \citenamefont {Persson},\
  and\ \citenamefont {Burgd\"orfer}}]{Arbo2010}%
  \BibitemOpen
  \bibfield  {author} {\bibinfo {author} {\bibfnamefont {D.~G.}\ \bibnamefont
  {Arb\'o}}, \bibinfo {author} {\bibfnamefont {K.~L.}\ \bibnamefont
  {Ishikawa}}, \bibinfo {author} {\bibfnamefont {K.}~\bibnamefont {Schiessl}},
  \bibinfo {author} {\bibfnamefont {E.}~\bibnamefont {Persson}}, \ and\
  \bibinfo {author} {\bibfnamefont {J.}~\bibnamefont {Burgd\"orfer}},\
  }\bibfield  {title} {\enquote {\bibinfo {title} {Intracycle and intercycle
  interferences in above-threshold ionization: The time grating},}\ }\href
  {\doibase 10.1103/PhysRevA.81.021403} {\bibfield  {journal} {\bibinfo
  {journal} {Phys. Rev. A}\ }\textbf {\bibinfo {volume} {81}},\ \bibinfo
  {pages} {021403(R)} (\bibinfo {year} {2010})}\BibitemShut {NoStop}%
\bibitem [{\citenamefont {Shvetsov-Shilovski}\ \emph
  {et~al.}(2016)\citenamefont {Shvetsov-Shilovski}, \citenamefont {Lein},
  \citenamefont {Madsen}, \citenamefont {R\"as\"anen}, \citenamefont {Lemell},
  \citenamefont {Burgd\"orfer}, \citenamefont {Arb\'o},\ and\ \citenamefont
  {T\ifmmode\mbox{\H{o}}\else\H{o}\fi{}k\'esi}}]{Shilovski2016}%
  \BibitemOpen
  \bibfield  {author} {\bibinfo {author} {\bibfnamefont {N.~I.}\ \bibnamefont
  {Shvetsov-Shilovski}}, \bibinfo {author} {\bibfnamefont {M.}~\bibnamefont
  {Lein}}, \bibinfo {author} {\bibfnamefont {L.~B.}\ \bibnamefont {Madsen}},
  \bibinfo {author} {\bibfnamefont {E.}~\bibnamefont {R\"as\"anen}}, \bibinfo
  {author} {\bibfnamefont {C.}~\bibnamefont {Lemell}}, \bibinfo {author}
  {\bibfnamefont {J.}~\bibnamefont {Burgd\"orfer}}, \bibinfo {author}
  {\bibfnamefont {D.~G.}\ \bibnamefont {Arb\'o}}, \ and\ \bibinfo {author}
  {\bibfnamefont {K.}~\bibnamefont
  {T\ifmmode\mbox{\H{o}}\else\H{o}\fi{}k\'esi}},\ }\bibfield  {title} {\enquote
  {\bibinfo {title} {Semiclassical two-step model for strong-field
  ionization},}\ }\href {\doibase 10.1103/PhysRevA.94.013415} {\bibfield
  {journal} {\bibinfo  {journal} {Phys. Rev. A}\ }\textbf {\bibinfo {volume}
  {94}},\ \bibinfo {pages} {013415} (\bibinfo {year} {2016})}\BibitemShut
  {NoStop}%
\bibitem [{\citenamefont {Eckle}\ \emph {et~al.}(2008)\citenamefont {Eckle},
  \citenamefont {Pfeiffer}, \citenamefont {Cirelli}, \citenamefont {Staudte},
  \citenamefont {D{\"o}rner}, \citenamefont {Muller}, \citenamefont
  {B{\"u}ttiker},\ and\ \citenamefont {Keller}}]{Eckle2008}%
  \BibitemOpen
  \bibfield  {author} {\bibinfo {author} {\bibfnamefont {P.}~\bibnamefont
  {Eckle}}, \bibinfo {author} {\bibfnamefont {A.~N.}\ \bibnamefont {Pfeiffer}},
  \bibinfo {author} {\bibfnamefont {C.}~\bibnamefont {Cirelli}}, \bibinfo
  {author} {\bibfnamefont {A.}~\bibnamefont {Staudte}}, \bibinfo {author}
  {\bibfnamefont {R.}~\bibnamefont {D{\"o}rner}}, \bibinfo {author}
  {\bibfnamefont {H.~G.}\ \bibnamefont {Muller}}, \bibinfo {author}
  {\bibfnamefont {M.}~\bibnamefont {B{\"u}ttiker}}, \ and\ \bibinfo {author}
  {\bibfnamefont {U.}~\bibnamefont {Keller}},\ }\bibfield  {title} {\enquote
  {\bibinfo {title} {Attosecond ionization and tunneling delay time
  measurements in helium},}\ }\href {\doibase 10.1126/science.1163439}
  {\bibfield  {journal} {\bibinfo  {journal} {Science}\ }\textbf {\bibinfo
  {volume} {322}},\ \bibinfo {pages} {1525} (\bibinfo {year}
  {2008})}\BibitemShut {NoStop}%
\bibitem [{\citenamefont {Corkum}\ \emph {et~al.}(1994)\citenamefont {Corkum},
  \citenamefont {Burnett},\ and\ \citenamefont {Ivanov}}]{Corkum_1994}%
  \BibitemOpen
  \bibfield  {author} {\bibinfo {author} {\bibfnamefont {P.~B.}\ \bibnamefont
  {Corkum}}, \bibinfo {author} {\bibfnamefont {N.~H.}\ \bibnamefont {Burnett}},
  \ and\ \bibinfo {author} {\bibfnamefont {M.~Y.}\ \bibnamefont {Ivanov}},\
  }\bibfield  {title} {\enquote {\bibinfo {title} {Subfemtosecond pulses},}\
  }\href {\doibase 10.1364/OL.19.001870} {\bibfield  {journal} {\bibinfo
  {journal} {Opt. Lett.}\ }\textbf {\bibinfo {volume} {19}},\ \bibinfo {pages}
  {1870} (\bibinfo {year} {1994})}\BibitemShut {NoStop}%
\bibitem [{\citenamefont {Milo\v{s}evi\'{c}}\ \emph {et~al.}(2006)\citenamefont
  {Milo\v{s}evi\'{c}}, \citenamefont {Paulus}, \citenamefont {Bauer},\ and\
  \citenamefont {Becker}}]{Milosevic2006}%
  \BibitemOpen
  \bibfield  {author} {\bibinfo {author} {\bibfnamefont {D.B.}\ \bibnamefont
  {Milo\v{s}evi\'{c}}}, \bibinfo {author} {\bibfnamefont {G.G.}\ \bibnamefont
  {Paulus}}, \bibinfo {author} {\bibfnamefont {D.}~\bibnamefont {Bauer}}, \
  and\ \bibinfo {author} {\bibfnamefont {W.}~\bibnamefont {Becker}},\
  }\bibfield  {title} {\enquote {\bibinfo {title} {Above-threshold ionization
  by few-cycle pulses},}\ }\href
  {http://stacks.iop.org/0953-4075/39/i=14/a=R01} {\bibfield  {journal}
  {\bibinfo  {journal} {J. Phys. B}\ }\textbf {\bibinfo {volume} {39}},\
  \bibinfo {pages} {R203} (\bibinfo {year} {2006})}\BibitemShut {NoStop}%
\bibitem [{\citenamefont {Popruzhenko}(2014)}]{Popruzhenko2014}%
  \BibitemOpen
  \bibfield  {author} {\bibinfo {author} {\bibfnamefont {S.~V.}\ \bibnamefont
  {Popruzhenko}},\ }\bibfield  {title} {\enquote {\bibinfo {title} {Keldysh
  theory of strong field ionization: history, applications, difficulties and
  perspectives},}\ }\href {http://stacks.iop.org/0953-4075/47/i=20/a=204001}
  {\bibfield  {journal} {\bibinfo  {journal} {J. Phys. B}\ }\textbf {\bibinfo
  {volume} {47}},\ \bibinfo {pages} {204001} (\bibinfo {year}
  {2014})}\BibitemShut {NoStop}%
\bibitem [{\citenamefont {Shvetsov-Shilovski}\ \emph
  {et~al.}(2012)\citenamefont {Shvetsov-Shilovski}, \citenamefont
  {Dimitrovski},\ and\ \citenamefont {Madsen}}]{Shilovski2012}%
  \BibitemOpen
  \bibfield  {author} {\bibinfo {author} {\bibfnamefont {N.~I.}\ \bibnamefont
  {Shvetsov-Shilovski}}, \bibinfo {author} {\bibfnamefont {D.}~\bibnamefont
  {Dimitrovski}}, \ and\ \bibinfo {author} {\bibfnamefont {L.~B.}\ \bibnamefont
  {Madsen}},\ }\bibfield  {title} {\enquote {\bibinfo {title} {Ionization in
  elliptically polarized pulses: Multielectron polarization effects and
  asymmetry of photoelectron momentum distributions},}\ }\href {\doibase
  10.1103/PhysRevA.85.023428} {\bibfield  {journal} {\bibinfo  {journal} {Phys.
  Rev. A}\ }\textbf {\bibinfo {volume} {85}},\ \bibinfo {pages} {023428}
  (\bibinfo {year} {2012})}\BibitemShut {NoStop}%
\bibitem [{\citenamefont {Eckart}(2019)}]{Eckart2019Diss}%
  \BibitemOpen
  \bibfield  {author} {\bibinfo {author} {\bibfnamefont {S.~G.}\ \bibnamefont
  {Eckart}},\ }\bibfield  {title} {\enquote {\bibinfo {title} {{Strong Field
  Ionization in Two-Color Fields}},}\ }\href@noop {} {\bibfield  {journal}
  {\bibinfo  {journal} {Ph.D. thesis, Johann Wolfgang Goethe-Universit\"at
  Frankfurt am Main}\ } (\bibinfo {year} {2019})}\BibitemShut {NoStop}%
\end{thebibliography}
\end{document}